\documentclass[sigconf]{acmart}

\AtBeginDocument{%
  \providecommand\BibTeX{{%
    \normalfont B\kern-0.5em{\scshape i\kern-0.25em b}\kern-0.8em\TeX}}}

\copyrightyear{2024}
\acmYear{2024}
\setcopyright{rightsretained}
\acmConference[CHI '24]{Proceedings of the CHI Conference on Human Factors in Computing Systems}{May 11--16, 2024}{Honolulu, HI, USA}
\acmBooktitle{Proceedings of the CHI Conference on Human Factors in Computing Systems (CHI '24), May 11--16, 2024, Honolulu, HI, USA}
\acmDOI{10.1145/3613904.3642847}
\acmISBN{979-8-4007-0330-0/24/05}

\usepackage{multicol}
\usepackage{multirow}
\usepackage{xspace}
\usepackage{xcolor}
\usepackage{makecell}
\usepackage{arydshln}
\usepackage{tabularray}
\usepackage{xcolor}
\usepackage[title]{appendix}

\newcommand{\sysname}{\textit{GenQuery}}
\definecolor{Issue1}{HTML}{AE3C60} % magenta
\definecolor{Issue2}{HTML}{DF473C} % pink
\definecolor{Issue3}{HTML}{F3C33C} % brick orange
\definecolor{Issue4}{HTML}{255E79} % purple
\definecolor{Issue5}{HTML}{267778} % emerald green
\definecolor{Issue6}{HTML}{82B4BB} % green

\begin{document}

% Title
\title{\sysname{}: Supporting Expressive Visual Search with Generative Models}

% Author
%%
\author{Kihoon Son}
\email{kihoon.son@kaist.ac.kr}
\affiliation{%
  \institution{School of Computing, KAIST}
  \city{Daejeon}
  \country{Republic of Korea}
}

\author{DaEun Choi}
\email{daeun.choi@kaist.ac.kr}
\affiliation{%
  \institution{School of Computing, KAIST}
  \city{Daejeon}
  \country{Republic of Korea}
}

\author{Tae Soo Kim}
\email{taesoo.kim@kaist.ac.kr}
\affiliation{%
  \institution{School of Computing, KAIST}
  \city{Daejeon}
  \country{Republic of Korea}
}

\author{Young-Ho Kim}
\email{yghokim@younghokim.net}
\affiliation{%
  \institution{NAVER AI Lab}
  \city{Seongnam}
  \country{Republic of Korea}
}

\author{Juho Kim}
\email{juhokim@kaist.ac.kr}
\affiliation{%
  \institution{School of Computing, KAIST}
  \city{Daejeon}
  \country{Republic of Korea}
}
% Author name shortcut
\renewcommand{\shortauthors}{Kihoon Son, DaEun Choi, Tae Soo Kim, Young-Ho Kim, and Juho Kim}

% Abstract
\begin{abstract}
Designers rely on visual search to explore and develop ideas in early design stages. However, designers can struggle to identify suitable text queries to initiate a search or to discover images for similarity-based search that can adequately express their intent. We propose ~\sysname{}, a novel system that integrates generative models into the visual search process. \sysname{} can automatically elaborate on users' queries and surface concrete search directions when users only have abstract ideas. To support precise expression of search intents, the system enables users to generatively modify images and use these in similarity-based search. In a comparative user study (N=16), designers felt that they could more accurately express their intents and find more satisfactory outcomes with ~\sysname{} compared to a tool without generative features. Furthermore, the unpredictability of generations allowed participants to uncover more diverse outcomes. By supporting both convergence and divergence, \sysname{} led to a more creative experience.
\end{abstract}

%%
%% The code below is generated by the tool at http://dl.acm.org/ccs.cfm.
%% Please copy and paste the code instead of the example below.
%%


% Keywords

% Teaser figure
\begin{teaserfigure}
  \centering
  \includegraphics[width=1.00\textwidth]{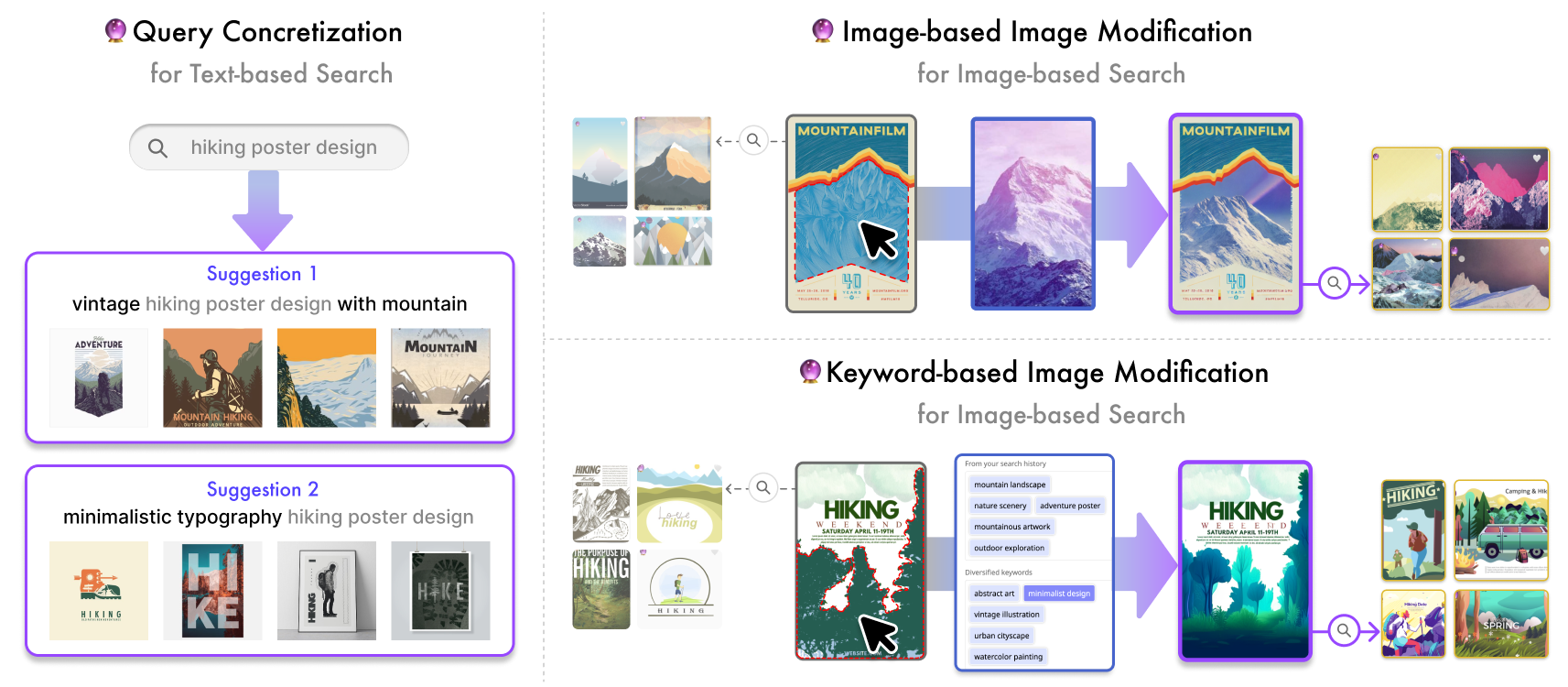}
  \caption{\textbf{Three main features of ~\sysname}: (1) Query concretization, (2) Image-based image modification, and (3) Keyword-based image modification feature. Query concretization concretizes the user's vague text query for the text-based search. Image-based image modification allows the user to select the area in an image (red dot line above) and to change the area of the image based on the reference image (purple mountain image). Keyword-based modification allows the user to change the selected area of an image (red dot line below) based on the keywords suggested from the user's search history. Both modification features' output could be utilized as an image-based search input and the results are changed due to the image modifications.}
  \Description{This figure illustrated three main features of ~\sysname. On the left, there is an illustration about Query concretization. Upon the search term "hiking poster design" is entered, two suggestions appear with a new search term and four search result images. Each suggestion is "vintage hiking poster design with mountain" and "minimalistic typography hiking poster design" respectively. On the right top, there is an illustration about Image-based image modification. There is a poster with a mountain, and the mouse clicks on the mountain part of it. Then, this part is modified with a new image with a photo of a purple-ish mountain, resulting in a new image of a poster with the purple mountain. The illustration also shows that the search results with a new modified image are different from the one searched by the original image. On the right bottom, there is an illustration about Keyword-based image modification. There is a poster with a forest scene, and the mouse clicks on the trees. Then, this part is modified with the keyword "minimalistic design", resulting in a new image of a poster with a flat blue-ish forest illustration. The illustration also shows that the search results with a new modified image are different from the one searched by the original image.}
  \label{fig:teaser}
\end{teaserfigure}

% Sections
\maketitle
\section{Introduction}
Visual search---searching and archiving diverse visual ideas---is an essential activity in the early ideation stage of the design process ~\cite{sharmin2009understanding}. In this process, designers find images close to their current ideas by entering a search query, assessing the search results, and setting new search directions for further exploration based on this assessment ~\cite{herring2009getting}. Beyond searching for images related to their current ideas (i.e., convergent thinking), designers also aim to uncover diverse and creative ideas through this visual search process (i.e., divergent thinking) ~\cite{goldschmidt2016linkographic}. This type of divergent exploration can help designers avoid fixating on specific ideas ~\cite{jansson1991design, herring2009getting} and drive the generation of more creative ideas~\cite{mougenot2008inspiration}. Various tools (e.g., Pinterest\footnote{https://co.pinterest.com/}, Behance\footnote{https://www.behance.net/}, or Dribbble\footnote{https://dribbble.com/}) support visual search with two methods: text-based ~\cite{tamura1984image} (i.e., inputting a text query to search for relevant images) and image-based search ~\cite{yeh2005picture, kang2021metamap} (i.e., inputting or clicking on an image to search for similar images).

However, designers can struggle to sufficiently express their intents (e.g., the type of designs they want to find or explore) through these search methods. For text-based search, it is challenging for designers to concretize their abstract thoughts into concrete search keywords~\cite{koch2020semanticcollage, mohian2022psdoodle}. On the other hand, image-based search partly addresses this challenge by allowing designers to search for designs by using other designs as queries~\cite{bunian2021vins, huang2019swire, ritchie2011d}. As these queried images can represent their search intent, the designer can search without having to put their thoughts into words. However, this type of search is typically limited to only retrieving designs that are \textit{similar} according to the tool, meaning that designers can neither explore divergently by searching for dissimilar designs nor designate specific aspects of designs for similarity search~\cite{mozaffari2022ganspiration}. Thus, designers can fail to effectively explore the design space as they do not know how to convey what they want or the tool does not allow them to do.

Through a formative study with eight designers, we probed deeper into designers' intents during visual search, their challenges in expressing these, and how they wish to express them. By asking participants to think-aloud while conducting a visual search, we observed that participants started with abstract intents and concretized these by repeatedly testing various search queries---which could be tedious and time-consuming. After concretizing a search query, participants explored around the design space by performing image-based search with designs that were most similar to their intents in the search results. However, although participants wanted to explore designs that were similar in terms of fine-grained aspects (e.g., mood, color, shape, or layout), the search tool used only considered the overall similarity of designs. For example, when converging on their search, participants wanted to search for designs that combined aspects from designs they already found and, when diverging, they wanted to search for designs that were similar to a found design in certain aspects but different in other aspects.

To address these problems, this work investigates how generative models---e.g., Large Language Models (LLMs) or Text-to-Image (T2I) models---could support intent expression during visual search. Specifically, we look at the ability of these models to take a rough idea and develop these into more concrete sketches.
For example, LLMs can effortlessly draft stories based on a starter sentence~\cite{chung2022talebrush, swanson2021story}, and T2I models can produce paintings and illustrations from a set of keywords~\cite{liu2022opal, chung2023promptpaint}.
In this sense, the generative models can be clear solutions for the aforementioned challenges in visual search: LLMs can expand on and concretize the user's vague intents, and T2I models can query images for the user when none of the search results match their intents.
While substantial work has highlighted the limitations of generative models in accurately executing users' intents~\cite{kim2023cells, wu2022ai, ko2023large, zhang2023adding}, we see an opportunity in integrating generative models as an intermediate layer in visual search tools: they can expand on users' intents and, as the outputs are leveraged as search inputs, the quality and accuracy of the generations are less consequential.
Furthermore, the unpredictability of these generative models can benefit the visual search process it could encourage spontaneous divergent explorations that can prevent fixation~\cite{jansson1991design}.

Therefore, we propose ~\sysname{} allows the users to concretize the abstract text query, express search intent through visuals, and diversify their search intent based on the search history. First, the user can concretize their vague search query through auto-complete suggestions with more specific search directions in text-based search (\autoref{fig:teaser} Query concretization). Second, the user can edit an image to generate an intent-aligned image as a search input through image-based image modification (\autoref{fig:teaser} Image-based modification). Third, the user can diversify their search intent through keyword-based image modification with the keywords suggested from the search history (e.g., saved visuals or inputted text queries). The modified image can also be used as a search input (\autoref{fig:teaser} Keyword-based modification).

To evaluate ~\sysname{}, we conducted a within-subjects study where 16 designers were asked to search for design ideas for two design tasks: ideation for a hiking club recruitment poster and an architecture exhibition poster. They were asked to save a minimum of five designs from the design ideation process using either ~\sysname{} or a baseline system similar to Pinterest, which allows only text- and image-based search without generation-based features.

Findings from our study demonstrated that ~\sysname{} allowed users to express their visual search intent easily and accurately compared to the baseline. Designers also felt the design ideas they discovered through ~\sysname{} were more diverse and creative significantly. Regarding the visual search pattern, the designer reduced the amount of text-based searches by ~\sysname{} by 71.2\% compared to the baseline. ~\sysname{} derived 34.4\% of \textit{search by generation} among the entire image-based search, and designers saved 35.8\% of the designs out of all the designs they saved through this new visual search pattern. As a result, designers were able to further develop their ideas in ~\sysname{} by roughly generating concepts and exploring similar results to the generated outputs. Based on these results, we discuss the degree of controllability over the generative outputs depending on how the user's search intern is concretized. We also discuss the various types of information for tracking the user's search intent more accurately in the visual search process.

This paper’s contributions are three-fold:
\begin{itemize}
    \item Findings from a formative study:  According to the convergent and divergent search stages, the study revealed how designers would express the search intent with what type of modalities and their related challenges.
    \item \sysname{}: Design and implementation of a visual search tool that incorporates generative models to enable designers to concretize and express their visual search intents effectively.
    \item Findings from a user study: The findings demonstrated that generative models can enhance user's accuracy of expression and flexibility in navigation during visual search, leading to more creative and diverse ideas.
\end{itemize}

\section{Related Work}

In our work, we aim to support designers' visual search process by aiding them in concretizing and expressing their search intents through generative outputs. To this end, we review related work in (1) the visual search process, (2) prior approaches to support search intent expression, and (3)generative models as an intermediate layer for intent expression.

\subsection{Visual Search for Early Design Ideation}

Visual search is an essential activity in designers' initial ideation process. During visual search, designers look for and explore existing designs relevant to their ideas and archive identified designs, typically by organizing them as mood boards~\cite{eckert2000sources}. This process enables designers to (1) generate, (2) develop, and (3) verify their ideas~\cite{eckert2000sources, bonnardel1999creativity}. In terms of generation, exploring diverse ideas can prevent design fixation ~\cite{jansson1991design}, where a designer may excessively focus only on a single idea. For development, as designers archive designs during visual search, they can then leverage these resources to develop their own creative concepts by taking inspiration from these ideas ~\cite{casakin1999expertise, goldschmidt2006variances, kang2018paragon} or to discuss ideas with other designers~\cite{lucero2012framing}. Finally, after developing a design direction, designers can quickly evaluate whether they want to further pursue this direction by observing actual design cases~\cite{herring2009getting}. Thus, visual search is a fundamental part of the design process, and assisting designers to search and explore references more effectively can enhance how they design.
Thus, due to the importance of the visual search process in the ideation and development of design ideas, our work aims to support by integrating generative models into the visual search process to further aid designers in concretizing, expressing, and diversifying their search intents.

\subsection{Supporting More Expressive Visual Search}

Due to the significance of visual search, there has been a wide array of tools that have been proposed to facilitate the process.
Typically, these tools support two main search methods: text-based search ~\cite{tamura1984image}, where a user inputs a text query to find relevant visuals, and image-based search ~\cite{yeh2005picture} (i.e., exemplar search ~\cite{kang2021metamap}), where a user inputs or selects an image to search for similar images. However, there are three main problems associated with these approaches. First, in text-based search, it can be challenging for designers to think of appropriate keywords that will return envisioned designs ~\cite{huang2019swire, bunian2021vins, fogarty2008cueflik, kovashka2012whittlesearch, mohian2022psdoodle, zhang2022tell, herring2009getting}. Second, while previous work has proposed more intuitive methods for users to input visuals instead of text queries (e.g., sketches~\cite{huang2019swire,bunian2021vins, mohian2022psdoodle, sain2021stylemeup}, wireframes~\cite{bunian2021vins} or example designs~\cite{ritchie2011d, zhang2022tell, fogarty2008cueflik}), these approaches depend on the user having a concrete idea of what type of design they are looking for. Third, as these search tools ~\cite{koch2020semanticcollage, huang2019swire, kovashka2012whittlesearch} focus on returning results that are similar to the input data but do not allow the user to control the degree of ``similarity'', users can struggle to look for results that align with their intents or to explore diverse ideas.

Several methods have been proposed that aim to enhance how designers can express their search intents by handling the limitations of text and example visuals as search inputs.
The Composing Text and Image to Image Retrieval (CTI-IR) method ~\cite{zhang2022tell} and Stylette~\cite{kim2022stylette} allows users to express their intents through both visuals \textit{and} natural language, and combines these modalities to retrieve more relevant images or suggestions. Leveraging a designer's previous actions, BIGexplore~\cite{son2022bigexplore} suggests images by interpreting the user's mouse actions in an exploratory search tool, and Kovacs et al.~\cite{kovacs2018context} propose a method that recommends images by interpreting designers' preferences from their previously used styles.
Furthermore, if the search results are unsatisfactory, WhittleSearch ~\cite{kovashka2012whittlesearch} modifies the search results through the user's natural language commands.
As a limitation of existing search tools is that they focus on returning similar results, prior work has also proposed approaches that instead suggest diverse designs. One thread of work focused on extracting the semantic meaning of images and then recommending images that are associated with that semantic meaning ~\cite{koch2020semanticcollage, kang2021metamap}. GANSpiration ~\cite{mozaffari2022ganspiration} uses a generative model to generate images of GUIs based on an input image randomly, and then searches for diverse GUIs based on their similarities with the generated images.

The aforementioned research has focused on supporting modalities that can help users to accurately express their search intents, methods that help users express their intent more easily, and approaches that allow users to explore diverse designs. However, no method has been proposed so far that can comprehensively address the problems in the visual search process to support the essential goal of the visual search process.
\label{section_rw2}

\subsection{Generative Models as a Channel for Intent Expression}
With the advancements in generative AI models (e.g., LLM or T2I), these models have been increasingly utilized across diverse tasks (e.g., writing~\cite{swanson2021story, yuan2022wordcraft, petridis2023anglekindling}, education~\cite{lee2023dapie}, and prototyping~\cite{jiang2022promptmaker}). 

Specifically, various researchers have investigated how to leverage the generative capabilities of these models to stimulate idea generation and facilitate creative thinking. For example, TaleBrush~\cite{chung2022talebrush} uses an LLM to quickly draft out a story based on a writer's sketch of a character's fortune. PopBlends~\cite{wang2023popblends} helps designers to ideate \textit{blends} between products and pop culture references by using an LLM to identify relevant ideas and their combinations. For screenplay writing, Dramatron~\cite{mirowski2023co} generates characters, descriptions, and actions based on a writer's outline to help them ideate on how to further progress and develop their screenplay. These work have demonstrated that LLMs can expand and elaborate on users' initial rough ideas, which can stimulate convergent and divergent thinking in creative tasks.

Researchers have actively conducted research on T2I models to provide visual outputs that are aligned with users' intent. As these models can produce high-quality images from a single prompt, various researchers have applied these to various domains, such as art ~\cite{john2023artinter}, graphic illustrations in news~\cite{liu2022opal}, 3D modeling ~\cite{liu20233dalle}, and fictional world-building~\cite{dang2023worldsmith}. However, controlling the generated outputs solely through text prompts can be challenging, which has led to the development of prompting guidelines ~\cite{liu2022design} and systems, like Promptify ~\cite{brade2023promptify} and RePrompt ~\cite{Wang_2023}, that support users in composing and selecting prompts for these models. Researchers have also explored novel techniques that allow for inputting different modalities into T2I models to enhance the controllability of their outputs further. These modalities include: reference images ~\cite{yang2023paint}, rough sketches ~\cite{voynov2023sketch}, canny edge ~\cite{zhang2023adding}, or object-segmented images ~\cite{goel2023pairdiffusion}. Beyond modalities, researchers have also proposed techniques that allow users to use these models for purposes beyond simple image generation. For example, there are techniques for stitching multiple objects or images~\cite{song2022objectstitch, sarukkai2023collage}, additional modification of generated output based on editing the inputted prompt ~\cite{hertz2022prompttoprompt, ge2023expressive, Mokady_2023_CVPR}, inpainting or infilling areas inside an image ~\cite{kandinsky2, saharia2022palette}, or extending an image by generating outside of its original boundaries ~\cite{saharia2022palette}. These threads of work have enabled the use of various modalities and interactions that can support more effective control of T2I models. In this work, we investigate how we can integrate these generative techniques into a visual search tool to help users more effectively and efficiently express their search intents.

\section{Formative Study}
To investigate the limitation of existing visual search tools (Section ~\ref{section_rw2}) within the actual search process and the user’s search intent to the challenges in detail, we conducted a formative study with eight designers using Pinterest\footnote{https://co.pinterest.com/}. Specifically, this study aims to observe the visual search process for ideation or brainstorming during early design stages. The main goal of this study is to understand the designers' overall patterns during visual search: 
\begin{itemize}
    \item When does the mode change between text-based search and image-based search?
    \item What is the search type among image- or text-based when they want to find a desired image or explore diverse images?
\end{itemize}
Next, we also aim to observe the user’s visual search intents when they faced the obstacles work Section ~\ref{section_rw2} through a think-aloud design and probe them on possible modalities that could support the expression of their intent. 

\subsection{Participants}
We recruited 8 designers (5 female and 3 male; D1-D8) in various design domains: architecture (D1 and D3), branding (D6 and D8), editorial design (D5 and D7), and web/mobile design (D2 and D4). All designers reported to have actively used visual search tools (Pinterest, Behance\footnote{https://www.behance.net/}, or Dribbbble\footnote{https://dribbble.com/}) during the early ideation stages of their design process.

\subsection{Procedure}
We asked participants first to perform a visual search task using a given tool (40 minutes), and then we conducted a semi-structured interview about their experience (20 minutes). We chose Pinterest as the search tool for the study as it is one of the most commonly used exemplar-based visual search tools ~\cite{kang2021metamap}. It allows users to input a text search query, select images to view similar examples, and filter search results based on high-level keywords.

For the visual search task, we provided three topics: startup landing page design, brochure design with an oriental painting aesthetic, and architectural poster design. Participants conducted the task with a think-aloud explaining their visual search intent (e.g., why they inputted a specific query or clicked on a visual). When participants struggled to fully express the intent verbally, we also asked them to explain themselves by referencing the images shown in the tool. For each search action that they performed, we also asked participants to explain what types of search results they were expecting or wanted to see before they actually looked at the results. After seeing the results, we then asked them to explain whether the results were satisfactory or not, and why this was the case. If the participant gave a non-specific or abstract explanation during the think-aloud, the authors prompted them to describe their intent more concretely through additional questions. 

After the task, we asked participants about how they conducted visual search in their actual design work, how well they felt they could express their search intents using the given tool, and whether the system could adequately understand their intent. We also asked participants about possible modalities that could help them express their intent better with visual search tools and about other additional supports that they wanted during the process.

\subsection{Visual Search Pattern, Problems, and Findings}
\label{section_formativefindings}
\begin{figure*}[!ht]
  \centering
  \includegraphics[width=1.00\textwidth]{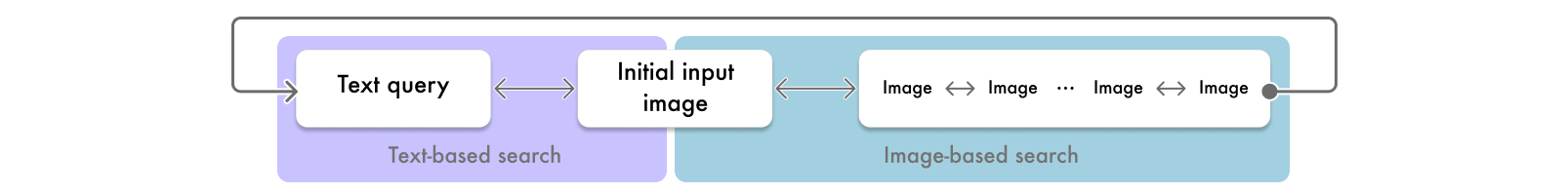}
  \caption{\textbf{The overall visual search process in an exemplar-based search tool:} Designers initiate their visual search with text-based search to find an image that they want to explore. When the designers find an image they would look for, they perform an image-based search by clicking the image. After seeing sufficient similar results with the initial input image (in the middle of an image-based search), they try to explore diverse images through the image-based search by clicking images that are different from the ones seen so far. In this process, when the designers feel they are continuously searching for similar images, they return to the text-based search to explore other images.}
  \Description{This figure shows an overall visual search process in an exemplar-based search tool. Designers initiate their visual search with text-based search to find an image that they want to explore. When the designers find an image they would look for, they perform an image-based search by clicking the image. After seeing sufficient similar results with the initial input image (in the middle of an image-based search), they try to explore diverse images through the image-based search by clicking images that are different from the ones seen so far. In this process, when the designers feel they are continuously searching for similar images, they return to the text-based search to explore other images.}
  \label{fig:visualSearchPattern}
\end{figure*}

The visual search process in an exemplar-based tool consists of a text-based search phase, where the designer inputs keywords to search for images, and an image-based search phase, where the designer clicks on images to explore other similar images. We observed that designers' visual search process follows a general structure (\autoref{fig:visualSearchPattern}).

\subsubsection{Iterative Search Query Editing Caused by Vague Initial Intent}
Designers initiate the visual search through the text-based search. All participants (D1-D8) initiated their search for the given task by first inputting vague text queries (e.g., landing page design or architecture poster). They explained that they search in this way because they do not know how to formulate their intent into search terms, and initially, their intent is not clearly defined. Participants said they usually start searching with a vague query and gradually refine it by adding one keyword at a time through repeated text-based searches. In this process, D3 mentioned that finding an appropriate search query can be challenging, as it takes a lot of time and there are instances where they cannot find the adequate query. Although the goal of the text-based search is to find an adequate image to start the image-based search, many participants (D1-D4 and D6-D7) face difficulties doing this with only the text-based search. Participants (D1-2 and D5) mentioned if they could get the concretized query and see the search results before inputting the query, the repetitive process of concretizing their initial search intent would be reduced.

\subsubsection{Difficulty in Finding an Intent-Aligned Visual in Image-based Search}
Participants initiated an image-based search from a visual found through the text-based search. Among the search results, participants chose the image most closely aligned with their search intent (i.e., most closely resembled what they were looking for). 

However, participants (D1, D3-D5, and D8) mentioned how it could be challenging to find an image that adequately expressed their intent. In many cases, most search results did not appeal to the participants (D1-D8). As participants had no other alternatives, they resorted to conducting multiple consecutive image-based searches where they clicked on the image that was closest to their intent until they were able to find images that represented their intent. While participants could not find one image that expressed their intent, we observed that several participants (D2, D4, and D6-D8) explained how multiple images could be combined to express their intent more accurately. D2 and D8 mentioned how, in their actual design work, they edited images in external tools and then used these edited images as references for image-based search.

As a reason for seeking the desired image as precisely as possible, participants (D1, D5, and D8) stated that confirming their envisioned visual elements and examining related search results help determine the next search direction. Ultimately, if they could not find the desired image during this process, they reverted to the text-based search, modifying the text query to initiate a new round of visual exploration.

\subsubsection{Diversifying Search Intent for Idea Exploration}
When the participants found sufficient images that resembled the visual aligned with their intent, they (D1-D8) started to search for alternatives that were different in terms of design elements (e.g., color, shape, composition, or layout). 
This reflects divergent thinking~\cite{frich2021how}, an essential aspect of the creative ideation process ~\cite{goldschmidt2016linkographic}. 

However, even when participants wanted to explore more diverse images, the current visual search tool predominantly only surfaced images with similar styles or overall color moods. When participants felt that the system did not provide diverse results, they would return to the first step of visual search (e.g., text-based search) and restart their exploration. 

Interestingly, when they started to look for diverse images, participants (D2-D4, D6, and D8) frequently employed a combination of abstract verbal terms and images to explain what kind of diverse search results they wanted (e.g., ``I would like to make this part more modern'' by D6). Regarding this, D4 mentioned how it was natural to use verbal explanations to express the types of diverse results due to the abstract nature of natural language. They believed that this ambiguity allowed them to explore a more diversified output.

\subsection{Design Goals}
Based on the findings, we defined three design goals for visual search interactions: 	
\begin{itemize}
    \item DG1: Provide recommendations for concretized search queries and each query's corresponding search results based on the initial query of the user.
    \item DG2: Support image-based image modification to clearly express the search intent of the user.
    \item DG3: Support keyword-based image modification to diversify the search intent of the user.
\end{itemize}

\section{\sysname{}}
We introduce ~\sysname{}, a system that leverages generative models to enhance the expression of visual search intent, thereby assisting users in their visual search process. ~\sysname{} provide text- and image-based search functionalities.
    
    \begin{figure*}[!ht]
      \centering
      \includegraphics[width=1.00\textwidth]{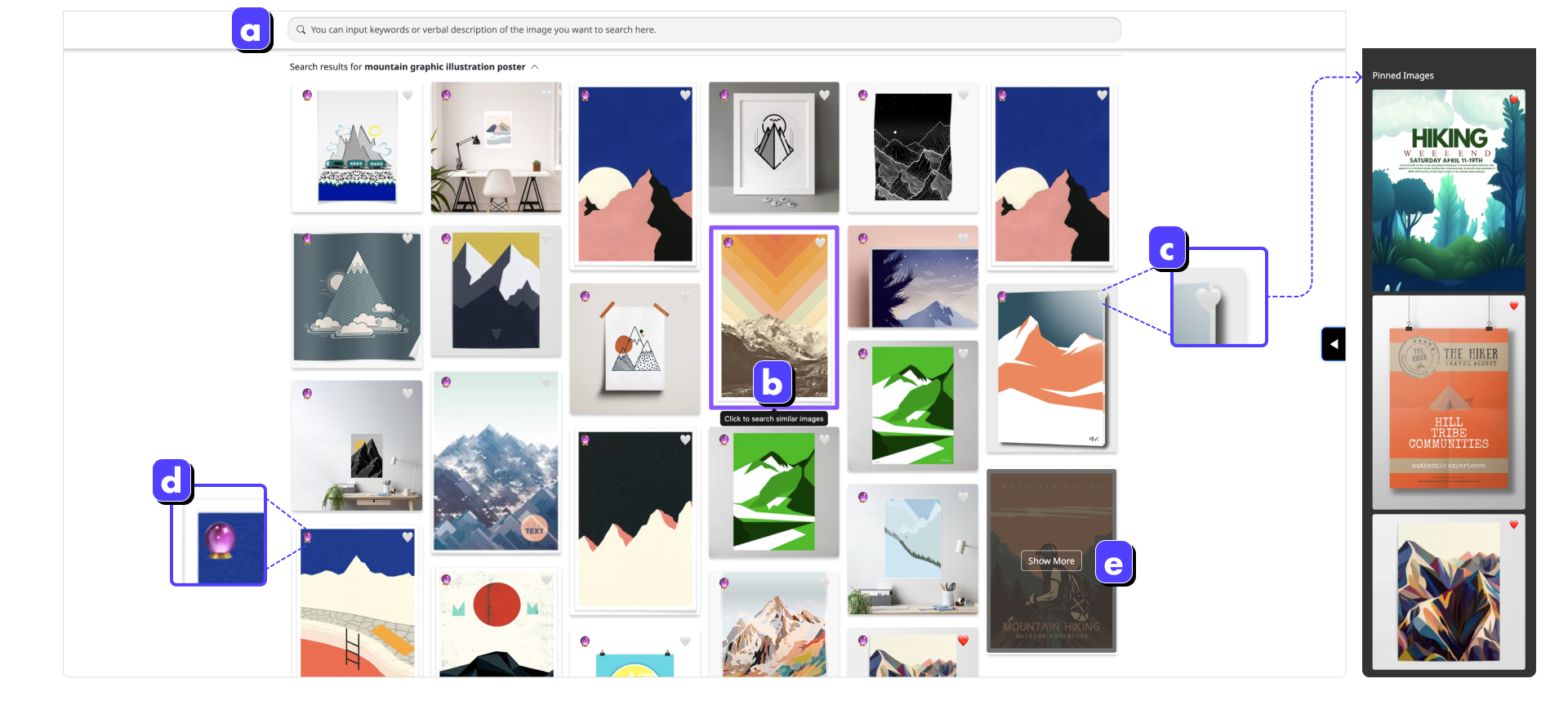}
      \caption{\textbf{Interface of ~\sysname{}}. \sysname{} shows the image search results as a gallery form. (a) Text prompt input box for text-based search: User can input a text description for the desired image here; (b) Clickable image for image-based search: An image in the gallery is clickable to provoke the image-based search. When the image is clicked, ~\sysname{} shows similar images to the clicked one at the bottom of the gallery; (c) Like button: The user can click the like button to save the design into the side panel; (d) Generation button: To edit one of the searched images for generating a new input, the user can click the marble emoji left top of image card. When the user clicks this button, the generation panel pops out below; (e) Show more button: This button is clicked when the user wants to see more search results}
      \Description{This figure shows a screenshot of an interface of ~\sysname{}. \sysname{} shows the image search results as a gallery form. On the top of the screen, there is a text prompt input box for text-based search. Users can input a text description for the desired image here. Below that, there is a gallery of search image results. All images are clickable, and when users click on them, the system provokes an image-based search to find similar images to the clicked one at the bottom of the gallery. For each image, there is a like button that users can click to save the design into the side panel. Also, there is a Generation button with a marble emoji in each image, where users can search images to generate a new input. When the user clicks this button, the generation panel pops out below. At the end of the gallery, there is a Show more button to see more search results.}
      \label{fig:desearch}
    \end{figure*}

In contrast to existing visual search systems, ~\sysname{} guides users in the early stages of the visual search process to help concretize their intent, particularly during the initial text-based search, where users might have vague text queries. \sysname{} provides concretized directions by adding related keywords for these vague queries, aiding the concrete formulation of the visual search intent. Next, during image-based searches, when users have a specific intent they wish to explore, ~\sysname{} generates visuals representing users' search intent by blending various images on a regional basis. Users can then base their searches on these mixed visuals to pinpoint their desired outcomes. Once the desired visual is found through image-based search, users can further explore diverse visual search results using keywords (e.g., ``A more minimalist style''). The system also considers users' search history to comprehend their intent and offers more diversified keywords. This enables users to modify their desired images in a broader spectrum, thus allowing for a richer visual search experience.

Users can utilize ~\sysname{} to concretize better (DG1), accurately express (DG2), and diversify (DG3) their search intent using generation-based interactions. Our system allows users to engage in an effective and inspiring visual search process, aligning with their design objectives.

\subsection{User Scenario}
This section describes how ~\sysname{} can assist visual search users in various situations through the process of \textit{Lily} using our system. \textit{Lily} is a student majoring in industrial design and she intends to design a poster to promote her hiking club to which she belongs. Before starting her design work, she decided to use \sysname{} to explore possible poster designs and to discover creative ideas.

\subsubsection{Concretize the Vague Search Query}
\begin{figure*}[!ht]
  \centering
  \includegraphics[width=1.00\textwidth]{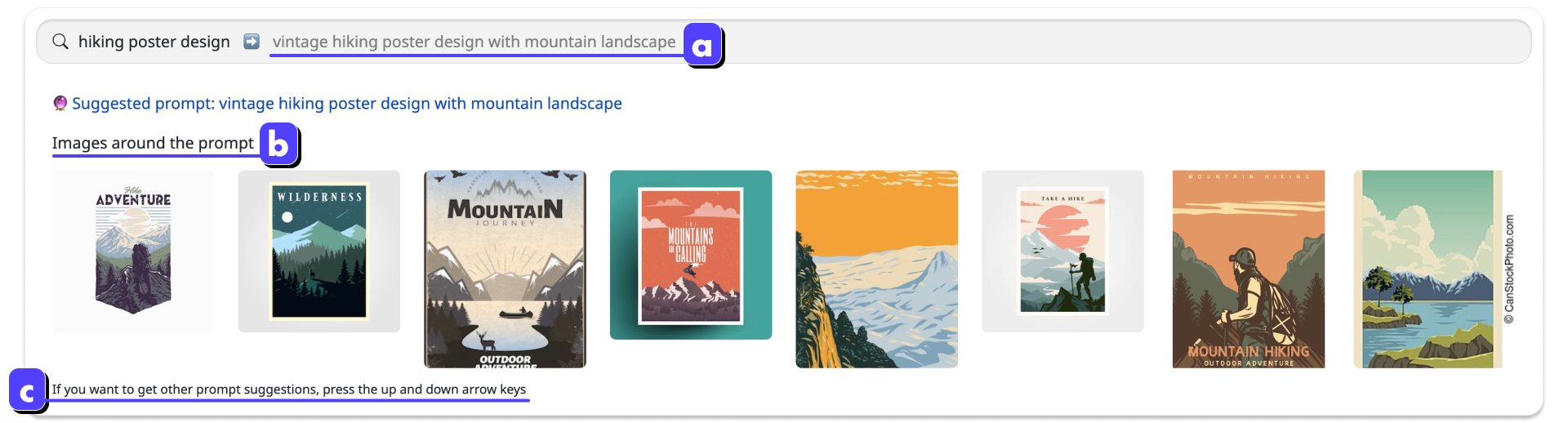}
  \caption{\textbf{Query concretization, which allows the user to concretize the user's initial abstract search query through LLM zero shot prompting}: (a) Suggested search query. User can swap their query by pressing the tab key; (b) Images searched by the suggested query are shown below in the search bar. Each image is clickable to process an image-based search; (c) ~\sysname{} provides five suggestions at a time, and the user can explore other suggestions by pressing the up and down arrow keys.}
  \Description{This figure shows a screenshot of the Query concretization feature of \sysname{}, which allows the user to concretize the user's initial abstract search query through LLM zero shot prompting. When the users give input ("hiking poster design") to the search bar, the suggested query ("vintage hiking poster design with mountain landscape") appears next to it, and users can use this query by pressing the tab key. Along with the suggested term, images that would be searched by the suggested query are shown below in the search bar. Each image is clickable to process an image-based search based on the clicked image. ~\sysname{} provides 5 suggestions at a time, and the user can explore other suggestions by pressing the up and down arrow keys.}
  \label{fig:genInteraction1}
\end{figure*}

Firstly, \textit{Lily} begins her search with the keyword \textit{``hiking poster design''} to explore various hiking poster designs. After entering the prompt in the top text search bar, ~\sysname{} provides \textit{Lily} with five suggestions on how to concretize her prompt (\autoref{fig:genInteraction1}-a), along with expected search results for each suggestion (\autoref{fig:genInteraction1}-b) in \autoref{fig:desearch}-a. \textit{Lily} probes the concretized prompts and corresponding search results by pressing the up or down arrow keys (\autoref{fig:genInteraction1}-c) to decide what sort of text prompt to input.

\subsubsection{Express Visual Search Intent Accurately through Image-based Image Modification}
\begin{figure*}[!ht]
  \centering
  \includegraphics[width=1.00\textwidth]{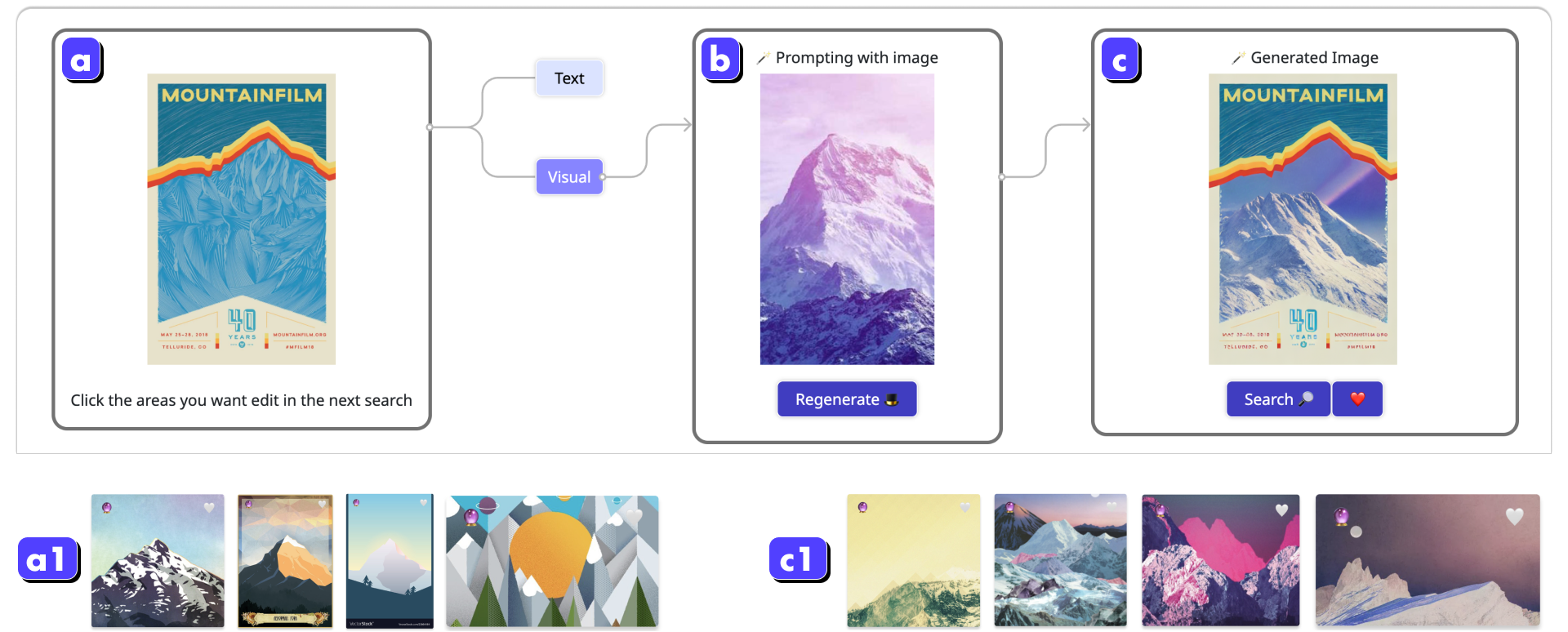}
  \caption{\textbf{Image-based image modification, which allows the user to express a clear search intent through reference-based editing and search}: (a) An image the user wants to modify. In this panel, the user can click or drag the areas in the image that he/she wants to modify. Currently, the blue ice mountain illustration has been selected; (b) An image the user wants to refer to in the editing process. The user can select a reference image from the search results or the saved design list; (c) The generation output. The user can do a regenerate or an image-based search with this generated result. The (a1) and (c1) show the difference in search results searched by the original image (a) and generated image (c).}
  \Description{This figure shows a screenshot of the Image-based image modification feature of \sysname{}, which allows the user to express a clear search intent through reference-based editing and search. In this figure, the user selects an image of a blue ice mountain illustration to modify, and a new panel comes out. The user can click or drag the areas in that image that they want to modify. Another image with a purple-colored mountain is shown next to it, which is the one that the user wants to refer to in the editing process. The generation output looks similar to the original one, but the mountain part is changed into a purple mountain with a sky background. The user can do a regenerate or an image-based search with this generated result. The screenshot also shows the difference in search results searched by the original image and the generated image.}
  \label{fig:genInteraction2}
\end{figure*}

While searching for images and navigating through an image gallery after inputting a text prompt, \textit{Lily} discovers a poster image she likes. She is happy with the overall composition and layout of the poster but doesn't quite like the mountain located in the middle of it. She wants posters depicting mountains of light purple color and finds another design representing the mountain design. Then, she wants to mix these two images to see if there are designs similar to the poster design she's envisioning. To this end, she clicks the generation button (\autoref{fig:desearch}-d), and then the generation panel is shown (\autoref{fig:genInteraction2}). She then selects the mountain area from the initially discovered poster image (\autoref{fig:genInteraction2}-a) and replaces this area with the mountain design above search results or in the side panel (\autoref{fig:genInteraction2}-b). Lastly, she clicks the generation button in \autoref{fig:genInteraction2}-b to see the generation output. She loves the generation result and proceeds with the image-based search using this generated image to find other similar designs (\autoref{fig:genInteraction2}-c). She finds the other images (\autoref{fig:genInteraction2}-c1) aligned with her intent.

\subsubsection{Diversify Visual Search Intent through Keyword-based Image Modification}

\begin{figure*}[!ht]
  \centering
  \includegraphics[width=1.00\textwidth]{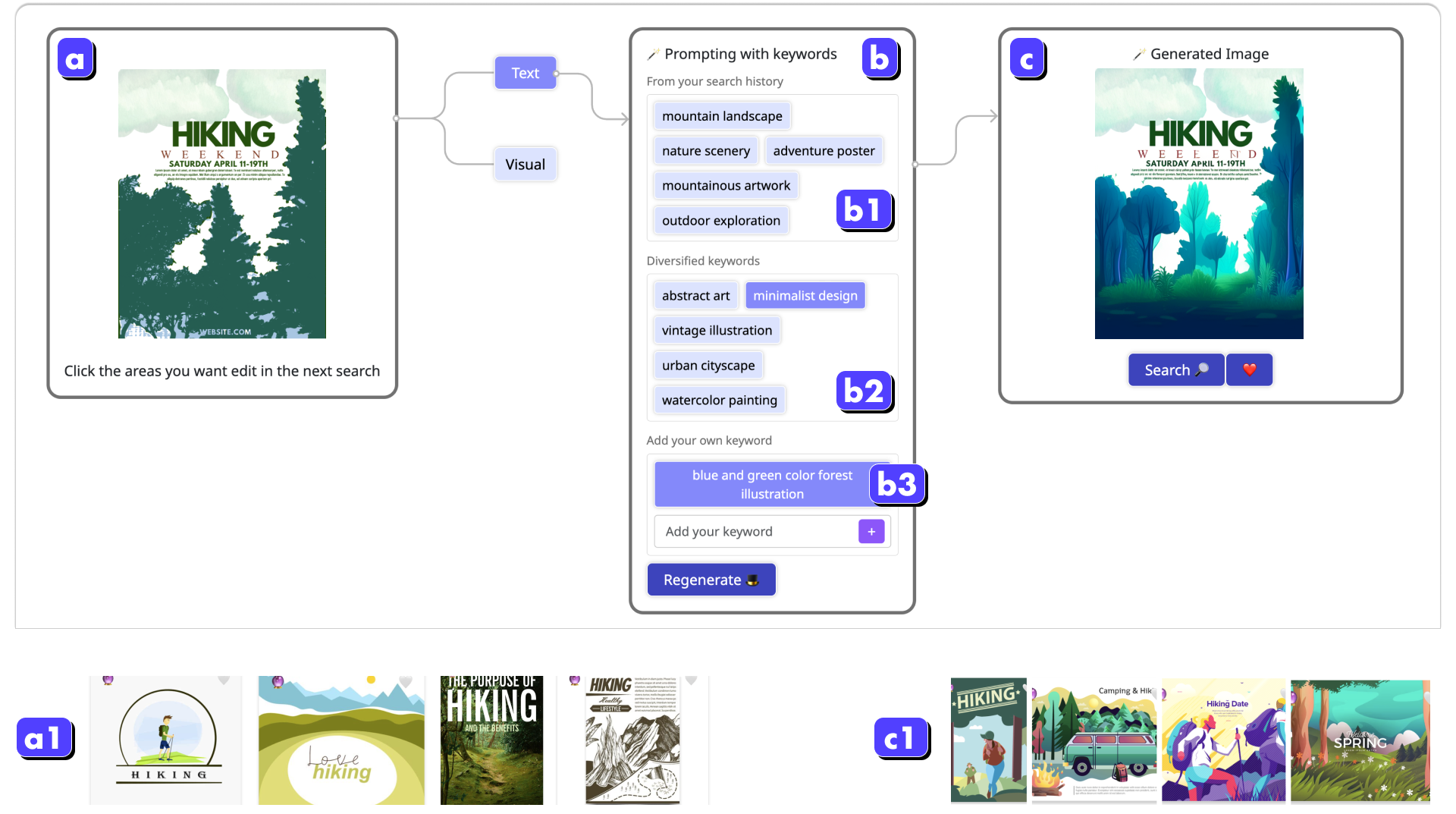}
  \caption{\textbf{Keyword-based image modification, which allows the user to diversify their search intent through keywords-based editing and search:} (a) An image the user wants to modify. Currently, the green forest illustration part has been selected; (b) Keywords suggestion panel based on the search history (e.g., search queries and saved image descriptions). ~\sysname{} suggest the keywords similar(b1) and new (b2) to previous search history. Also, the user can make their own keyword in b3; (c) The generation output. The user can do a regenerate or an image-based search with this generated result. The (a1) and (c1) show the difference in search results searched by the original image (a) and generated image (c).}
  \Description{This figure shows a screenshot of the Keyword-based image modification feature of \sysname{}, which allows the user to diversify their search intent through keywords-based editing and search. In this figure, the user selects an image of a green forest illustration to modify, and a new panel comes out. The user can click or drag the areas in that image that they want to modify. A small panel is shown next to it, and there are three sections of keywords. The first section shows the keywords derived from the user's search history, such as "mountain landscape". The second section shows the diversified keywords, such as "abstract art". The third section provides users a custom input, so the user enters "blue and green color forest illustration". The user selects "minimalistic design" from the diversified keywords part, and "blue and green color forest illustration" from the custom input section. The generation output looks similar to the original one, but the forest part is changed into a flat illustration of green and blue forest. The user can do a regenerate or an image-based search with this generated result. The screenshot also shows the difference in search results searched by the original image and the generated image.}
  \label{fig:genInteraction3}
\end{figure*}

After a certain amount of image searches about mountains, \textit{Lily} begins to worry that she has been too focused on the mountain landscape for the poster design. She believes that the view a person would see while hiking would also be appropriate for a club poster image. So, she starts to look for poster images that feature forest scenes, and she finds a poster. (\autoref{fig:genInteraction3}-a). However, the forest in this poster is only expressed in a single green color, so she decides to modify it.

However, since she is unsure how to modify it, she is considering editing the image with text keywords to see the possible modification directions of this design. So she clicks the text button in the generation panel. Looking at the keyword suggestions, she can find some keywords related to her previous search history (\autoref{fig:genInteraction3}-b1). Also, she discovers that she hasn't looked at many designs related to \textit{``minimalism''} yet (\autoref{fig:genInteraction3}-b2). Thus, she selects the tree area (\autoref{fig:genInteraction3}-a) and inputs the keyword \textit{``blue and green color forest illustration''} along with the suggested keyword \textit{``minimalist''} (\autoref{fig:genInteraction3}-b3). After seeing the result, she adores this design. Then, she starts a new image-based search in this direction using this newly generated image and finds search results related to the new direction (\autoref{fig:genInteraction3}-c1).

\subsubsection{Basic Visual Search Functions in ~\sysname{}}
Fundamentally, ~\sysname{} offers functionalities similar to existing visual search tools (e.g., Pinterest), such as searching for visuals with a text prompt (\autoref{fig:desearch}-a), the operation to click an image to find similar images (\autoref{fig:desearch}-b), and the feature to save desired images and archive them in the right-side panel (\autoref{fig:desearch}-c). By clicking the \textbf{Show more} button, more results can be viewed from the current search results (\autoref{fig:desearch}-e). In the system, the search results are all provided to the user in a format stacked under the most recent search results.

\subsection{Technical Details}
In this section, we introduce more technical details of \sysname{}. We specifically focus on the three key functionalities and implementation details.

\subsubsection{Query Concretization}
~\sysname{} provides five concretized queries based on the text query entered by the user in ~\autoref{fig:desearch}-(a), about one second later. To ensure a smooth user experience with ~\sysname{}, it was critical to deliver suggestions at near-real-time speed; hence, we used the gpt-3.5-turbo-16k-0613 LLM model. To deliver concretized queries as output, we applied zero-shot prompting with the [Current Search Query] entered by the user, following the prompting instruction in the ~\autoref{appendix:a}. For more accurate results, we prompted the LLM to explain the reasoning behind the concretized queries. Using these concretized queries, ~\sysname{} performs image-based searches and presents the top eight results as shown in ~\autoref{fig:genInteraction1}-(b).

\subsubsection{Image-based Image Modification}
Upon the user clicking on ~\autoref{fig:desearch}-(d), the panel in ~\autoref{fig:genInteraction2} is displayed. In this process, the server, leveraging SAM ~\cite{Kirillov2023SegmentA}, segments the clicked image into individual objects and sends this to the front-end, enabling users to make selections within the image by area. When the user selects an image to refer to and presses the Generation button in ~\autoref{fig:genInteraction2}-(b), the server constructs a mask corresponding to the area selected in ~\autoref{fig:genInteraction2}-(a). Subsequently, utilizing the mask image, the original image, and the reference image, a new image output is generated through the PaintByExample model ~\cite{yang2023paint}. We decided to use this model considering the type of modality utilized by the model and its fast generation time (3-4 sec). The resulting creation is displayed in ~\autoref{fig:genInteraction2}-(c).
~\label{section_generation2tech}

\subsubsection{Keyword-based Image Modification}
In the process outlined in \S\ref{section_generation2tech}, after selecting the area of the image, the user can also press the Text button instead of the Visual to see~\autoref{fig:genInteraction3}-(b). To aid the user in modifying the image based on keywords, we performed zero-shot prompting using the same LLM model with the query suggestions. To elaborate, we used the description of the image the user wants to modify ([Description of Current Image]), the last five search queries ([Search Query History]), and descriptions of the last five saved images ([Descriptions of Saved Images]) in our prompting. Through this, we have made the LLM generate keywords that are similar to those that have been searched so far, as well as different keywords. To achieve better suggestion output, we instruct LLM to predict the user's search intent and suggest keywords in the prompt (See ~\autoref{appendix:b}). Also, we prompt the model to include the reasons for the predicted search intent in its answer. \sysname{} displays these generated keywords in ~\autoref{fig:genInteraction3}-(b). Once the user has selected the keywords to modify the image and clicked the Generation button, the Kandinsky2.2 ~\cite{kandinsky2} diffusion model is used to generate a new image based on the original image, mask image, and selected keywords. As in ~\S\ref{section_generation2tech}, we chose to use this model in ~\autoref{fig:genInteraction3} considering the modality used by the diffusion model and the generation speed (2-3 seconds).

\subsubsection{Implementation Details}
~\sysname{} is a web-based system, which is composed of a ReactJS front-end and a Python Flask server as the back-end. For the image dataset and basic search functions (e.g., text-based search and image-based search), we used the \texttt{clip-retrieval} library~\cite{beaumont2022clipretrieval} on the hosted API provided by the library. Given an input query or image, the library processes the input into an embedding and then queries similar images from the LAION-5B dataset~\cite{schuhmann2022laion5b} through the API. We used a machine with an AMD Ryzen 9 5900X 12-Core Processor and NVIDIA GeForce RTX 3090 to implement ~\sysname{}.

\section{Evaluation}
To investigate the impact of the Query concretization (\autoref{fig:genInteraction1}), Image-based (\autoref{fig:genInteraction2}), and Keyword-based image modification (\autoref{fig:genInteraction3}) of ~\sysname{}, we conducted a within-subjects study with 16 designers. This study aims to comprehensively investigate how the visual search experience and designers' search patterns change when the generative support is incorporated. The three features proposed for ~\sysname{} are built upon traditional visual search features like standard text- and image-based search. For comparison, we establish a baseline that retains these traditional features and excludes generation support, distinguishing it from other approaches like generation-focused interfaces such as Opal~\cite{liu2022opal} or multi-dimensional search-based methods such as MetaMap~\cite{kang2021metamap}. We believe that comparing with this standard-like baseline will more effectively showcase the generation's impact on the visual search process. From the study's main objective, the baseline followed the design of \sysname{}: it supported the same image-based and text-based search features and allowed for saving designs. The baseline supports the same main features of famous visual search tools like Pinterest or Dribbble.

In the study, participants performed two different visual search tasks using both ~\sysname{} and the baseline, and saved designs of interest during their explorations. Through this user study, we focused on answering the following research question:
\begin{itemize}
    \item RQ1: Can ~\sysname{} support the user's satisfactory visual search process?
    \item RQ2: How do the methods of intent expression of ~\sysname{} change the user's visual search patterns compared to the visual search process of baseline?
    \item RQ3: What are the positive and negative effects of the generation process in the visual search process?
\end{itemize}

\subsection{Participants}
For the study, we recruited a total of sixteen participants (Age Mean = 27.19, Age SD = 4.09, Female = 10, and Male = 6). We recruited participants who met the following conditions: 1) majoring in design-related majors or currently working in design, and 2) have extensive experience using exemplar-based tools to conduct visual search during the initial stages of the design process. In order to recruit participants from more diverse design domains, we recruited people through online advertisements and word-of-mouth. As a result, we recruited participants in industrial design (six participants), architecture and interior design (four participants), and graphic design (five participants).

\subsection{Tasks and Procedure}

The study was conducted both in-person and online. For participants who could not participate in person, we conducted the study online using Zoom ~\footnote{https://zoom.us/}. The tasks given in the study involved performing a visual search for poster design ideas for 20 minutes. Here, we instructed the participants that the goal of visual search is design ideation, similar to the early stage of the design process. Thus, we explained to the participants that the final goal of this study is to save at least five visuals as design ideas during the process. This task is based on a visual search process of the early design phase for ideation, which was identified from the formative study. Although there are various ways to design ideation, we specifically oriented the participants to focus on conducting the ideation process only through the visual search with the provided search tools. Additionally, we selected the poster design task. The tasks given to participants were 1) a poster design for a hiking club advertisement and 2) a poster design for an architecture exhibition.

The total study time was 1 hour and 30 minutes, which started with an introduction to the study and asking participants for their informed consent. Then, for each task, participants were provided with a 5-minute tutorial of the tool they would be using and proceeded to perform the task for 20 minutes. After each task, participants responded to a survey for 5 minutes. To avoid ordering effects, both the conditions used in the tasks and the tasks themselves were counterbalanced. Participants were provided with a 5-minute break between tasks. After participants completed both tasks, we conducted semi-structured interviews for 20 minutes to investigate the differences in participants' experiences between the two conditions. As compensation, we provided participants with 45,000 KRW (approximately 34 USD).

\subsection{Measures}
\label{section_measures}
For measures, the surveys conducted after each task included questions that asked participants to rate on a 7-point Likert scale their satisfaction with the saved designs, the quality of these designs ~\cite{kang2021metamap}, their satisfaction with the search process, and Behavior Intention ~\cite{venkatesh2008tam}. The survey also included questions that asked participants to rate how well they could express their search intent, along with questions related to the Creativity Support Index (CSI) ~\cite{cherry2014csi}. Additionally, questions from the NASA-TLX questionnaire ~\cite{hart1988nasa} were also included in the survey to investigate the potential additional workload that may arise from the generative features. In the case of standardized measures like CSI and NASA-TLX, we conducted a paired t-test. Except for these measures, we conducted a Shapiro-Wilk test to assess the normality of the data. Subsequently, we used a paired t-test for parametric data, while a Wilcoxon signed-rank test was applied for non-parametric data.

In the in-depth interviews, the interview questions focused on what differences participants perceived between their search processes when they had or did not have the generative features. Also, we asked participants on whether there were differences in how they expressed their intents and, if so, what impact it had on their search processes. We also inquired about participants' search patterns by asking them about when they switched between types of search (e.g., image-based vs text-based), when they continued to use the same type of search, and what pattern did they perform more frequently for each condition and why. Furthermore, we also asked participants whether they performed any new patterns when they had the generative features of ~\sysname{}.

Additionally, we analyzed participants' interaction logs to measure the number of designs that they saved. To investigate how the overall search processes differed between conditions, we measured the number of text- and image-based search actions that were performed. Furthermore, we also measured how often different search patterns occurred: text-based search to another text-based search (\textbf{T-T}), text-based to image-based (\textbf{T-I}), image-based to image-based (\textbf{I-I}), and image-based to text-based (\textbf{I-T}). Lastly, we analyzed patterns related to the generative feature, such as the proportion of cases where the generation results were saved, the proportion of searches performed with the generated results, and the proportion of results saved after a generation-based search.
% \labelparametriceasures

\section{Results}

In the study, we observed that participants used the generative features of ~\sysname{} to more accurately express their visual search intent and, as a result, obtain more satisfactory search results. Participants expressed to continue using ~\sysname{} than baseline significantly. Additionally, we found that, due to the generative features, participants could find more diverse and creative design ideas while performing fewer text- and image-based searches. Also, we found a new visual search pattern through ~\sysname{}, ~\textit{Search by Generation}, which means an image-based search through the image- or keyword-based modified image output.

\begin{figure*}[!ht]
  \centering
  \includegraphics[width=1.00\textwidth]{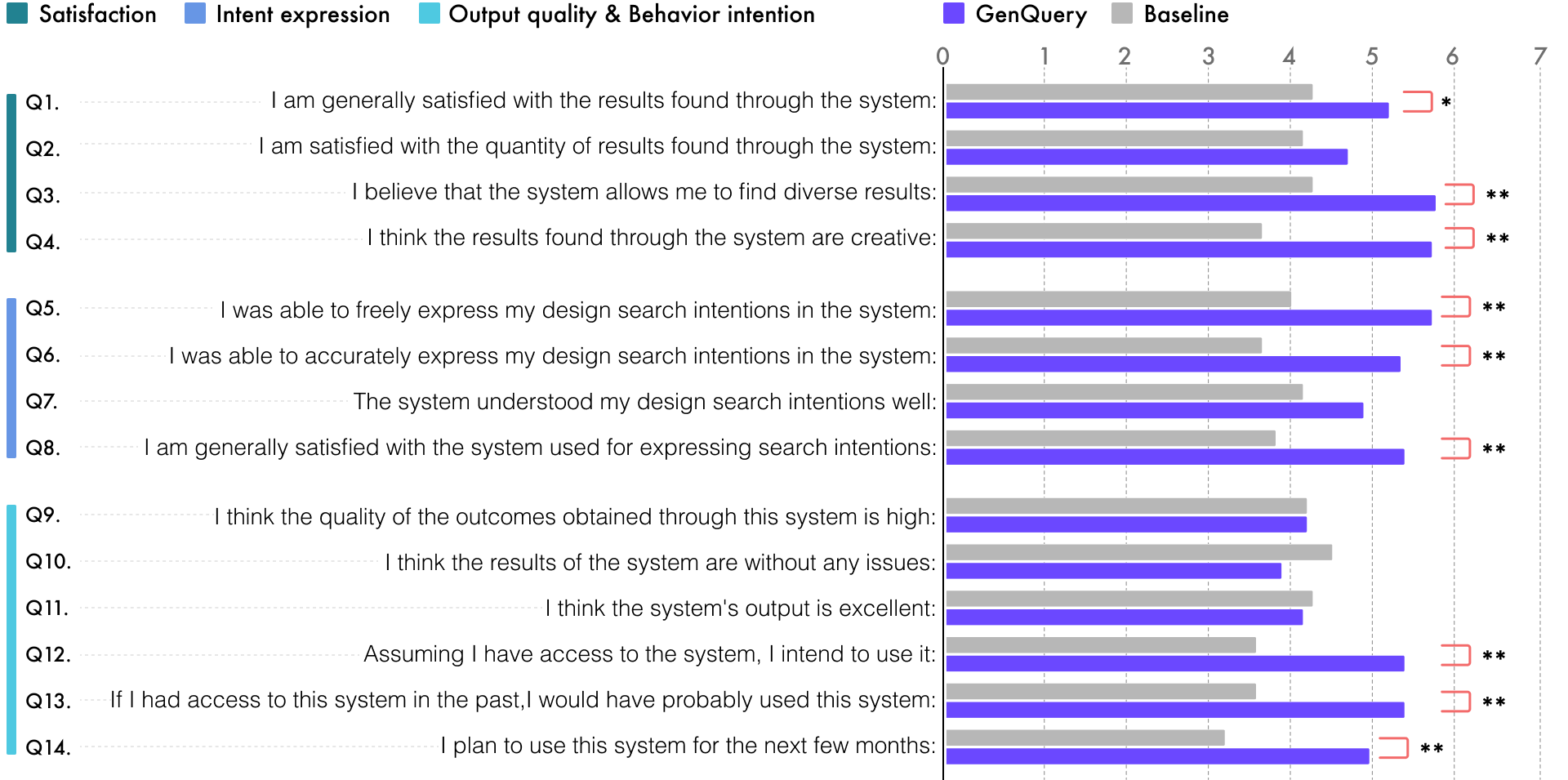}
  \caption{\textbf{Survey results in the user study.} The survey asked participants to rate their Satisfaction ~\cite{kang2021metamap}, Intent Expression, and Output Quality \& Behavior intention ~\cite{venkatesh2008tam} on a 7-point Likert scale. \textasteriskcentered{}: p<0.05 and \textasteriskcentered{}\textasteriskcentered{}: p<0.01.}
  \Description{This graph shows survey results in the user study. There are four questions regarding user satisfaction, and \sysname{} is shown to be significantly better in three statements, including "I am generally satisfied with the results found through the system", "I believe that the system allows me to find diverse results", and "I think the results found through the system are creative". There are another four questions about intent expression, and \sysname{} is shown to be significantly better in three statements, including "I was able to freely express my design search intentions in the system", "I was able to accurately express my design search intentions in the system", and "I am generally satisfied with the system used for expressing search intentions". There are also six questions about the output quality and behavior intention, and \sysname{} is shown to be significantly better in three statements, including "Assuming I have access to the system, I intend to use it", "If I had access to this system in the past, I would have probably used this system", and "I plan to use this system for the next few months"}
  \label{fig:survey}
\end{figure*}

\subsection{RQ1: High Satisfaction with Visual Search Process Using the Generative Features}
Participants rated their satisfaction with the search results to be significantly higher when using ~\sysname{} (\autoref{fig:survey}; Q1, Mean\_diff=0.94, p<0.05). In particular, participants felt significantly higher satisfaction in terms of the diversity (\autoref{fig:survey}; Q3, Mean\_diff=1.50, p<0.01) and creativity (\autoref{fig:survey}; Q3, Mean\_diff=1.69, p<0.001) of the discovered ideas. Additionally, participants willingness to use each tool in the future was significantly higher when they used ~\sysname{} (\autoref{fig:survey}; Q14, Mean\_diff=1.75, p<0.01). P4 mentioned that the process of exploring designs became more enjoyable due to the integration of generative steps. 
Most participants stated in interviews that they preferred the designs found through ~\sysname{} (13/16). P5 said that this was because ``\textit{the generative feature seemed to consistently suggest new directions during the visual search process where one can continue exploring unconsciously}''. Particularly, when using the baseline, participants (P4, P8, and P12) described their process to mostly involve repeatedly choosing an image from the search results that are closest to what they want, not exploring various design ideas. On the other hand, ~\sysname{} generation allowed them to search for ideas diversely; participants showed a strong desire to use ~\sysname{} in their design process (\autoref{fig:survey}; Q12, Mean\_diff=1.81, p<0.01).

Another reason for participants' higher satisfaction with the search results in ~\sysname{} was that the ideas found through the system included more advanced design ideas than their initial ideas. Participants attributed this to the different ways in which the generative features supported the search process. First, participants (P1, P7, P11, and P13) described how the query concretization ~\autoref{fig:genInteraction1} allowed them to dive into more quickly concrete design ideas compared to when they started with abstract search queries. Second, participants (P2, P4-P6, P8, P12, and P14) mentioned the image- and keyword-based image modification (\autoref{fig:genInteraction2} and \autoref{fig:genInteraction3}) helped them to think about the next design search direction by actually observing search results similar to the modification output. In particular, P4 said, ``\textit{In the initial design process, it is essential to save the ideas and merge those again to review whether the idea is good or not. However, in this tool, you can simply and quickly try the design you want, so the design process supported by ~\sysname{} seemed to cover both the initial and middle stages of design.}'' This was a view echoed by many participants (P2, P4-P5, P8, P12). In summary, ~\sysname{} assisted designers in more quickly and easily navigating the developed design ideas during the visual search process, leading to higher satisfaction levels.

On the other hand, there was no significant difference in the quantity of designs found between \sysname{} and the baseline (In \autoref{fig:survey} Q2, Mean\_diff=0.56, p>0.05; Baseline\_mean = 12.44, Baseline\_SD = 5.67; The average number of saved designs ~\sysname{}\_mean = 9.75, \sysname{}\_SD = 4.23, p>0.05). Regarding this, P6 said that there were cases where she had to repeatedly modify the prompt or image used in the generative features because the generated results did not satisfy her. Furthermore, P6 mentioned that the generation process led her to repeatedly modify an image unknowingly until the system created a result that she wanted. In the visual search process using \sysname{}, the images generated by the modification process account for 45.8\% (SD=28.6\%) of the total saved ideas. P10 said \textit{``In the conventional search process, even if undesired results appeared, I could easily ignore them because there is no opportunity to modify the image. But the generation process of ~\sysname{} made me immersed in the generation process itself.''} As there were quite a few instances where the quality of the generated results was low or the generation went in an unpredictable direction, participants showed the opinion that there was not much difference in the quality of the entire results obtained from the system (\autoref{fig:survey}; Q9, Mean\_diff=0.00, p>0.05). Specifically, P5 stated that there were cases when he was not very satisfied just by looking at the results generated by the system, and about 30-40\% of the generated results were not satisfactory. All participants who expressed dissatisfaction during the creation process had specific images they wanted to generate, but the generative model failed to meet their needs.

In summary of the findings for RQ 1, ~\sysname{} provided great satisfaction overall, especially regarding the diversity and creativity of the final identified design products. However, it did not provide great satisfaction in terms of the quality of the generated product itself and the overall quantity of the identified products.

\label{section_result_RQ1}

\subsection{RQ2: Findings of New Visual Search Patterns with A More Efficiency}
Participants were able to express their visual search intent more accurately through ~\sysname{}, and as a result, they conducted a more efficient visual search process. As shown in ~\autoref{fig:survey}, participants could express their visual search intent more freely (\autoref{fig:survey}; Q5, Mean\_diff=1.69, p<0.01) and accurately (\autoref{fig:survey}; Q6, Mean\_diff=1.69, p<0.01) in ~\sysname{}. Overall, they were highly satisfied with this (\autoref{fig:survey}; Q8, Mean\_diff=1.56, p<0.01). However, as mentioned in Section ~\ref{section_result_RQ1} about the low controllability of the generation process, there was no significant difference in the survey asking whether the system understood their search intent well (\autoref{fig:survey}; Q7, Mean\_diff=0.75, p>0.05). P3 explained that he could elaborate their search intent through ~\sysname{} compared to the baseline, and though it was not perfect, it was somewhat possible to express it. Particularly, P3 stated that the overall visual search process seemed to be more efficient due to the generation process.

\begin{figure*}[ht]
  \centering
  \includegraphics[width=1.00\textwidth]{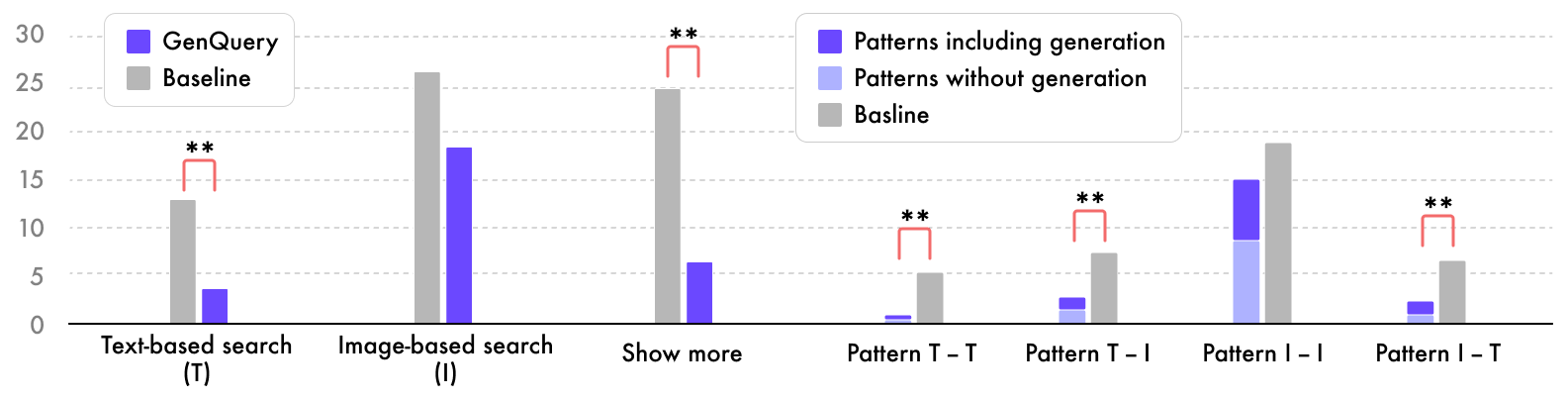}
  \caption{\textbf{Visual search pattern analysis results.} The left side graph shows the difference between ~\sysname{} and baseline in terms of text-based search (\textbf{T}), image-based search (\textbf{I}), and \textbf{Show more} button-clicking action. The right side graph shows the difference between ~\sysname{} and baseline in terms of search pattern of \textbf{T-T}, \textbf{T-I}, \textbf{I-I}, and \textbf{I-T}.}
  \label{fig:pattern}
  \Description{This graph shows visual search pattern analysis results. \sysname{} is shown to be significantly less in terms of text-based search and "Show more" button click action. Also, there is significantly fewer instances in the search pattern of T-T (text-based search after text-based search), T-I (image-based search after text-based search), and I-T (text-based search after image-based search)}
\end{figure*}

In fact, analyzing the logs of the participants in both systems (\autoref{fig:pattern}), when using the baseline, participants significantly performed more Text-based search (\textbf{T}) (\sysname{}\_mean = 3.69, Baseline\_mean = 12.81, p<0.001) and \textbf{Show more} (\sysname{}\_mean = 6.50, Baseline\_mean = 24.44, p<0.01) actions. The \textbf{Show more} button, which is pressed to see more related search results from the search results, is equivalent to scrolling down on Pinterest. In Image-based search (\textbf{I}), although there's not a significant difference (\sysname{}\_mean = 18.38, Baseline\_mean = 26.31, p>0.05), participants who used ~\sysname{} definitely took fewer actions (\autoref{fig:pattern}). Despite the difference in the number of actions, there was no significant difference in the number of final saved design ideas (Baseline\_mean = 12.44, Baseline\_SD = 5.67, ~\sysname{}\_mean = 9.75, \sysname{}\_SD = 4.23, p>0.05). According to the participants (P1-P3, P5, P8, P12, and P14), the generative feature that recommends a text query prevented unnecessary query editing processes, and the process of generating and searching for images reduced many actions performed to find the desired image within the image.

In order to analyze how the search pattern in the visual search process has changed specifically new ways of search intent expression, we also performed a pattern analysis of Text-based Search (\textbf{T}) and Image-based Search (\textbf{I}) as explained in ~\S\ref{section_measures}. In the interviews session of the study, we asked about the meaning of \textbf{T-T}, \textbf{T-I}, \textbf{I-I}, and \textbf{I-T} in the visual search process, and as a result, we were able to summarize the meanings of each pattern as follows. The patterns are similar to what we found in the ~\S\ref{section_formativefindings}.
\begin{itemize}
    \item \textbf{T-T}: Pattern performed when unable to decide on the desired design area to start a visual search
    \item \textbf{T-I}: Pattern performed when at least some of the desired visuals are included in the search results
    \item \textbf{I-I}: Pattern performed when the desired image is currently visible or when wanting a result that is somewhat slightly different than the results found so far
    \item \textbf{I-T}: Pattern performed when wanting to refresh the search process because it seems only to find similar images
\end{itemize}

As seen in ~\autoref{fig:pattern}, apart from \textbf{I-I}, participants were able to drastically reduce the count of the other three types of patterns when using ~\sysname{} (T-T\_diff = -4.94; T-I\_diff = -6.00; I-T\_diff = -5.63; All p<0.01). Interview results revealed that many participants (P3, P6-P7, and P10-P16) felt they particularly performed \textbf{T-T} and \textbf{I-T} patterns often in the baseline. For this reason, P12 mentioned, ``\textit{I feel like I've performed these two patterns quite often as I frequently face difficulties to find the appropriate search term or when the visuals I'm currently looking for feel not so novel.}'' In contrast, when using ~\sysname{}, they reported to have often performed the \textbf{I-I} pattern (P3, P5, P9-P12, and P14-P15). P14 elaborated, ``\textit{Finding the desired image was relatively easy, and once I found this image, I wanted to look for a slightly different image. The process of modifying the image for a generation was very suitable for this type of pattern.}'' Notably, P3 said that \textit{``It was possible to generate by referencing other images in the two features of modifying the image, and also, the keywords showed what kind of things I have mainly searched for so far (\autoref{fig:genInteraction3}), that helped me understand in what direction I should explore at that point.''} In other words, ~\sysname{} had a pronounced impact on the overall visual search pattern, and the \textbf{I-I} pattern, which became the core of the visual search, had a significant role.

\begin{figure*}[ht]
  \centering
  \includegraphics[width=1.00\textwidth]{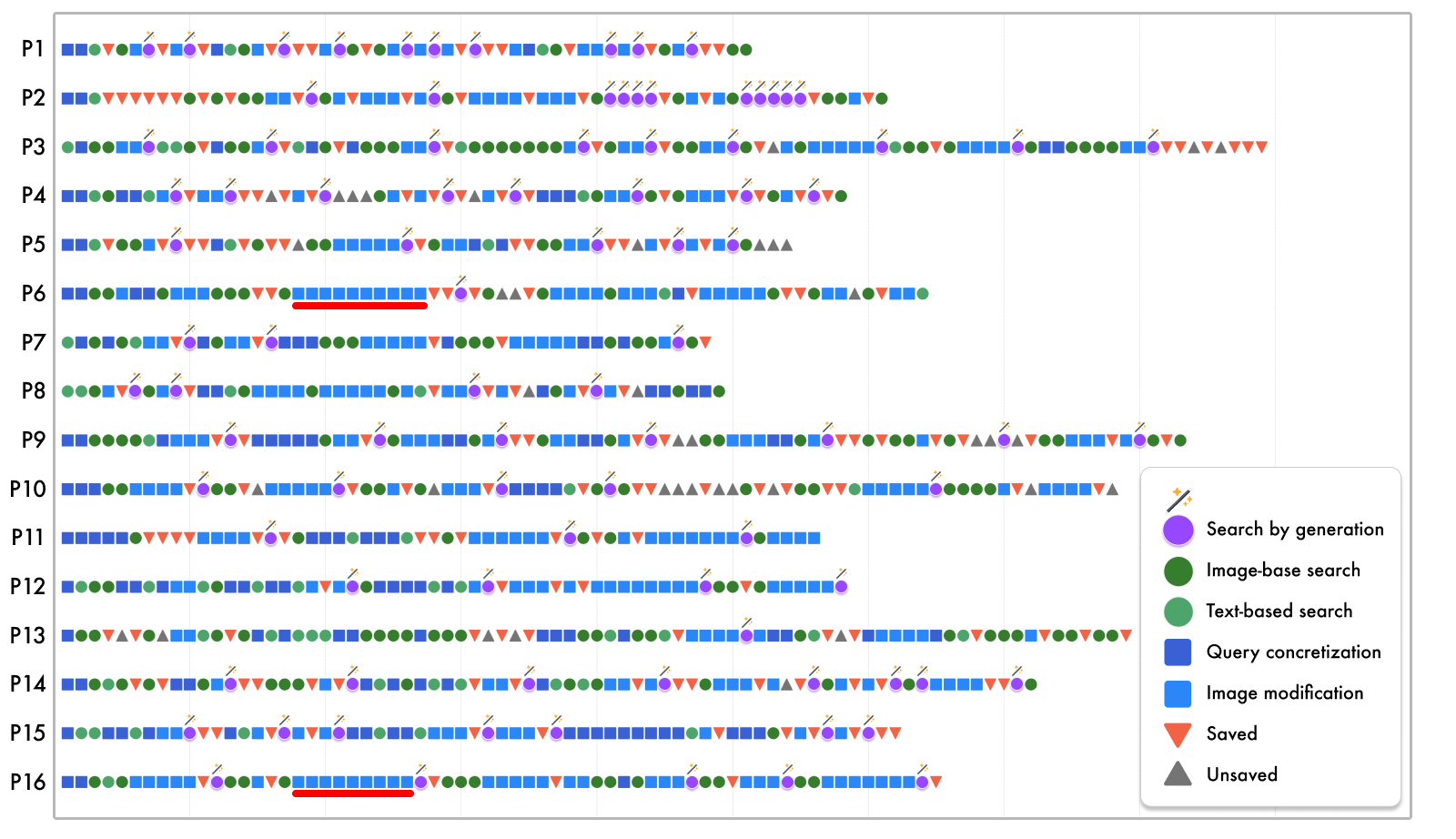}
  \caption{\textbf{All participants' action log visualization in ~\sysname{} including \textit{Search by Generation}, Image-based search, Tex-based search, Image modification (\autoref{fig:genInteraction2} and \autoref{fig:genInteraction3}), Query concretization (\autoref{fig:genInteraction1}), Saved, and Unsaved actions:} Generation search means an image-based search by the generated output. The two red lines indicate the unsatisfactory generation cases because the participants were deeply engrossed in the generation process without searching. The actual images generated from these processes are illustrated in ~\autoref{fig:generationProcess} as well.}
  \Description{This plot shows action logs of all participants using \sysname{}. Search by Generation, Image-based search, text-based search, image modification, query concretiation, saved, and unsaved actions are shown in the plot. The legend shows on the right side. We can see the image modification action is frequently included during the image-based search. Also, the plot shows there are text-based searches. The two redlines mean that the unsatisfactory experience of participants P6 and P16 was caused by the unexpected output from the generative models.}
  \label{fig:logproxy}
\end{figure*}

\begin{figure*}[!ht]
  \centering
  \includegraphics[width=1.00\textwidth]{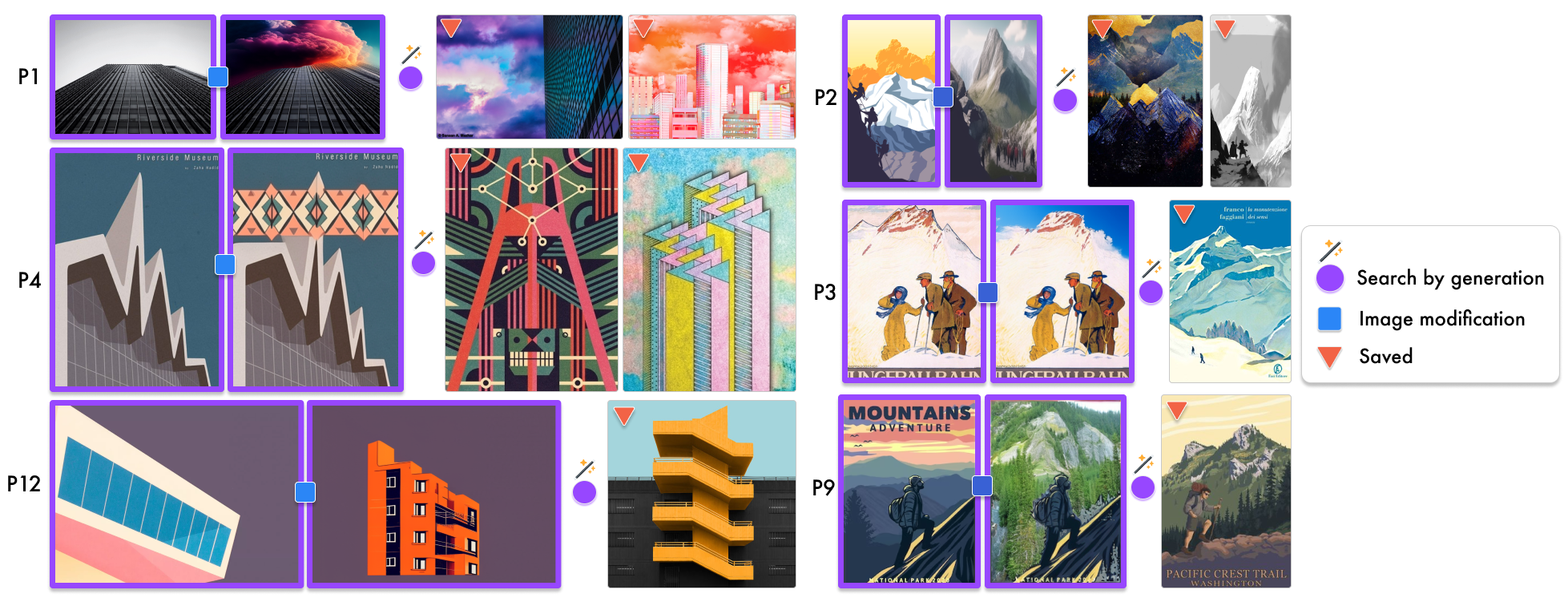}
  \caption{\textbf{The actual examples of saving design ideas through \textit{Search by Generation} in ~\sysname{}.} The left side is the case of task 1, and the right side is the example of task 2.}
  \Description{This figure shows actual examples of saving design ideas through Search by Generation in ~\sysname{}. The left side is the case of task 1, and the right side is the example of task 2. A total of six participants' example of Search by Generation is depicted. Each image group shows the original image, modification image, and search result images by the modification image.}
  \label{fig:exampleGenerationSearch}
\end{figure*}

Furthermore, we observed that due to the generative feature, participants could generate a design expressing the direction they wanted to explore, search with it, and consequently save the images they wanted. We named this search pattern as \textit{Search by Generation}. As can be seen in ~\autoref{fig:pattern}, the four patterns on the right are represented in ~\sysname{} divided into cases with and without image generation actions included between them. When using ~\sysname{}, image generation actions are included between the actions in \textbf{T-T}, \textbf{T-I}, \textbf{I-I}, and \textbf{I-T} at rates of 66.7\% (SD=42.4\%), 46.7\% (SD=37.5\%), 47.7\% (SD=22.5\%), and 62.9\% (SD=44.6\%) respectively. Specifically, the pattern of \textit{Search by Generation} is 34.4\% (SD=20.1\%, Min=3.8\%, Max=87.5\%) among the image-based search. P6, who performed the lowest amount of \textit{Search by Generation}, mentioned, ``As I got more immersed in the process of generating the final result, I couldn't think much about using it for search.'' In contrast, P15, who performed the most generation searches, stated, ``The feature to modify and search really allowed me to express my ideas more accurately, so I found myself conducting a lot of those searches. It was a feature that I had always felt was necessary while using platforms like Pinterest.'' Also, the ~\autoref{fig:exampleGenerationSearch} illustrates the examples of \textit{Search by Generation} we observed in the study. While there may be variations among participants, we did observe this search pattern frequently results at various points throughout the visual search process ~\autoref{fig:logproxy}. When using ~\sysname{}, the ratio of searching with a newly generated image and saving a design within the results was also calculated. This value was 35.8\% (SD=21.3\%) of the total saved images. In other words, Participants saved around 36\% images through the \textit{Search by Generation} among the total amount of saved designs.
Search through generation

Based on the findings, we have demonstrated how the visual search process has changed with the generative features of ~\sysname{}. The generative features made the search process more efficient by helping to express the precise visual search intent, making it possible to find a similar quantity of results more satisfactorily with fewer search actions.

\label{section_result_RQ2}

\subsection{RQ3: Strength and Weakness of Unexpectedness of Generative Model and Creativity}
The unexpectedness provided by the generative model, especially the T2I model generating images, offered both positive and negative aspects to participants during the visual search process. 

\begin{table*}[!ht]
    \small
    \begin{tabular}{@{}rrcccccc@{}}
    \toprule
    \multicolumn{1}{l}{}                                                                             &               & \multicolumn{2}{c}{\textbf{\sysname{}}} & \multicolumn{2}{c}{\textbf{Baseline}} & \multicolumn{2}{c}{Statistics} \\ \cmidrule(l){3-8} 
    \multicolumn{1}{l}{}                                                                             &               & mean         & std          & mean          & std          & p-value               & Sig.         \\ \midrule
    \multirow{6}{*}{NASA-TLX~\cite{hart1988nasa}}                                                                        
     & Mental        & 2.56         & 1.36         & 2.69          & 1.82         & 0.86            & -            \\
     & Physical      & 1.25         & 0.77         & 2.06          & 2.02         & 0.06            & +            \\
     & Temporal      & 1.63         & 0.81         & 2.38          & 2.06         & 0.19            & -            \\
     & Effort        & 3.69         & 1.70         & 3.31          & 1.58         & 0.53            & -            \\
     & Performance   & 5.00         & 1.15         & 4.88          & 0.89         & 0.74            & -            \\
     & Frustration   & 2.00         & 1.59         & 2.69          & 2.06         & 0.17            & -            \\ \bottomrule
    
    \multirow{5}{*}{Creativity Support Index (CSI) ~\cite{cherry2014csi}}                                                                        
     & Enjoyment        & 8.19         & 1.88         & 4.19          & 0.93         & 0.00            & \textasteriskcentered{}\textasteriskcentered{}\textasteriskcentered{}            \\
     & Exploration      & 8.28         & 1.13         & 3.53          & 1.22         & 0.00            & \textasteriskcentered{}\textasteriskcentered{}\textasteriskcentered{}            \\
     & Expressivness      & 8.25         & 1.46         & 3.75          & 1.34         & 0.00            & \textasteriskcentered{}\textasteriskcentered{}\textasteriskcentered{}            \\
     & Immersion        & 7.03         & 2.93         & 4.63          & 1.34         & 0.01            & \textasteriskcentered{}\textasteriskcentered{}            \\
     & Results Worth Effort   & 7.41         & 1.75         & 4.88          & 1.07         & 0.00            & \textasteriskcentered{}\textasteriskcentered{}\textasteriskcentered{}            \\
     & Collaboration   &     ---     &     ---     &     ---      &    ---      &      ---       &  --- \\
     \bottomrule
    \end{tabular}
    \caption{\textbf{The results of NATA-TLX (\cite{hart1988nasa}; 7-Likert Scale) and Creativity Support Index (\cite{cherry2014csi}; 10-Likert Scale) survey}: Since ~\sysname{} dosen't support the collaboration with other designers, we excluded the Collaboration related questions (we assigned the weight value as zero) questions in the final calculation of CSI score. -: p > .100, +: .050 < p < .100, \textasteriskcentered{}: p < .050, \textasteriskcentered{}\textasteriskcentered{}: p < .010, \textasteriskcentered{}\textasteriskcentered{}\textasteriskcentered{}: p < .001}
    \Description{This table shows the ressult of NASA-TLX and Creativity Support Index in both \sysname{} and baseline condition. There are no significance between the two condition in NASA-TLX, while \sysname{} is shown to be significantly better in all indices of Creativity Support Index compared to the baseline.}
    \label{table:survey}
\end{table*}

Firstly, on the negative side, participants thought the quality of the results generated by the model was relatively poorer than the actual designs. However, what made the search experience worse was the low controllability to generate desired images, as confirmed in the interview response from P10. P10 said using ~\sysname{} was very challenging, citing the difficulty of controlling the text-to-image model as the reason. He explained that, especially when he had images he wanted to create in mind, he found it more difficult to accept the generated result and seemed to get deeply immersed in the generation process due to repetitive prompting processes (e.g., changing the reference image or keywords). Due to this low controllability of the model, three participants were unsatisfied with ~\sysname{}.

However, as seen in ~\autoref{table:survey}, this generation process did not cause an additional burden. On the other hand, many participants (P1-P2, P4-P5, P7, P11-P13, and P16) reported being led to contemplate unexpected exploration paths due to the model's unforeseen image generations. In interviews, several participants (P1, P4, P12, and P16) stated they were able to plan their next search process by witnessing the generation in an unexpected direction. Related to this, P16 commented, ``\textit{Even though the generation output was strange at times, I became more curious about the search outcome through it. And when I tried to generate, I got a result that I couldn't imagine in my head, so I actually searched with it.}'' P16 mentioned that this process was interesting as it felt like evaluating the results generated by the AI. 

Furthermore, the unexpectedness resulting from low controllability served as a starting point for users to engage in various attempts. P2 remarked, ``\textit{I was initially puzzled seeing the generation results, but shortly after, a new idea popped up looking at it},'' and further added, ``\textit{Through this generation process, it was possible to consider up to ten ideas in the search process that usually involves contemplating only three.}'' Even, P3 described experiencing a sense that the part they had to think about diminished when using ~\sysname{}. ~\sysname{} consistently proposed new directions within the range set by the user, and as a result, participants using ~\sysname{} responded in the survey that they had overwhelmingly creative experiences compared to the baseline (\autoref{table:survey}, CSI).

We discovered the strengths that can be exhibited in terms of design creativity when leveraging the incompleteness of T2I models in the visual search process. Although some participants felt discomfort regarding this aspect, ultimately, our study results show that such incompleteness could be sufficiently utilized as an interaction in the visual search process, and more broadly, in the visual search process that requires diverse stimulation. This turned the perceived flaw of the T2I model into a feature beneficial to the creative process.

\section{Discussion}

\subsection{Generation Process in Visual Search: The Controllability on Output Depends on the Specification Level of Intent}
\begin{figure*}[ht]
  \centering
  \includegraphics[width=1.00\textwidth]{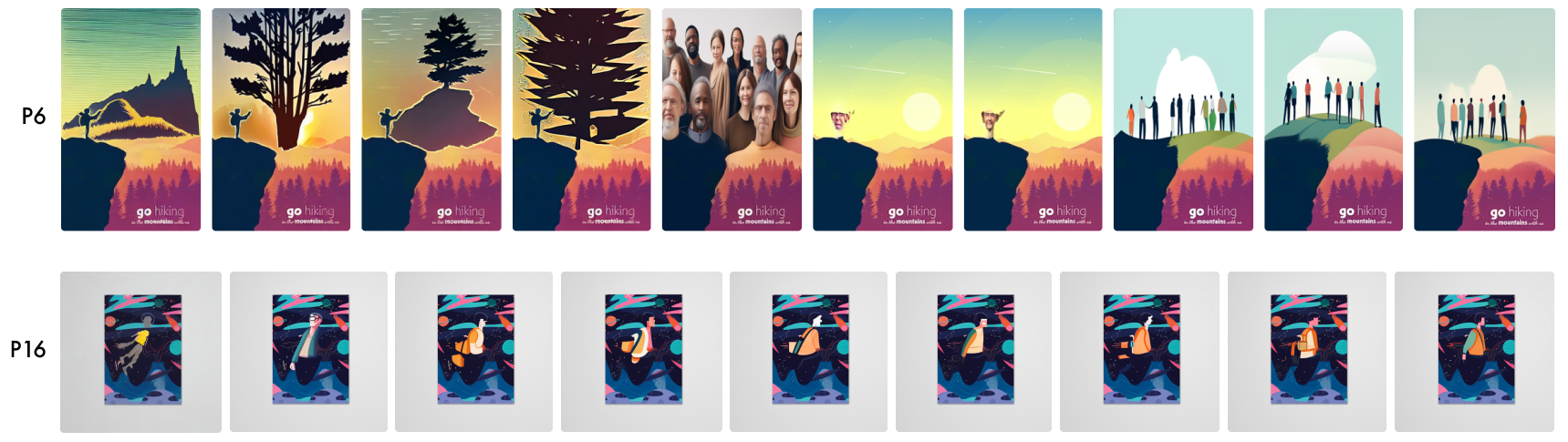}
  \caption{\textbf{Examples of engrossing into the generation process of P6 and P16}: P6 tried to generate the background of the image with trees and people standing on the hill. P16 tried to generate a hiking person illustration in the center of the image iteratively. Both felt unsatisfactory and explained that they had wanted to make the image that they imagined in the interview session.}
  \Description{This figure shows examples of engrossing into the generation process of P6 and P16. P6 tried to generate the background of the image with trees and people standing on the hill. P16 tried to generate a hiking person illustration in the center of the image iteratively. Both felt unsatisfactory and explained that they had wanted to make the image that they imagined in the interview session.}
  \label{fig:generationProcess}
\end{figure*}

In our study, the user's level of concreteness of the search intent influenced the experience of participants with the generation results of \sysname{}. We observed that participants became more engrossed in the generation process when the image they wanted to see was clearly defined, as shown in ~\autoref{fig:generationProcess}. Particularly, when the generated results contained low-quality details, participants became fixated on the generation results (\autoref{fig:generationProcess}). P6 stated, ``\textit{I wanted to draw many people on the hill, and I had a specific image in my mind. However, the generated results were different from what I expected, so I kept pressing the generate button.}'' P16 said, ``\textit{The legs of the person were not appearing, so I kept generating. I wanted to draw at least the legs.}'' As participants tried to directly generate the desired image without \textit{search by generation}, with unexpected outcomes resulting in a poorer user experience, and ultimately, we observed required higher controllability in this case. P6 and P16 mentioned that previewing what kind of design could be discovered through these generated results would prevent getting caught up in the generation process.

On the other hand, from participants' comments (P8 and P14), we observed that the lower controllability on generative models works well when the user's search intent was not concrete or when the fine details of output were not critical (e.g., generating backgrounds or incorporating artistic elements). With this degree of intent, P8 was willing to use the generated results for visual search, even if they differed from their expectations. In the case of generation related to abstract styles rather than specific objects, participants engaged more in the search process without getting deeply involved in the image modification process. Interestingly, in these cases, many participants often reconsidered their search direction based on the generated results. 

In summary, we observed the generative models' output provides different user experiences based on the status of the search intent of users. A design lesson for future work is the level of controllability over the generation output would be differentiated according to the concretization level of the user's search intent.

\subsection{Generation Process as Design Prototyping}
\begin{figure*}[ht]
  \centering
  \includegraphics[width=1.00\textwidth]{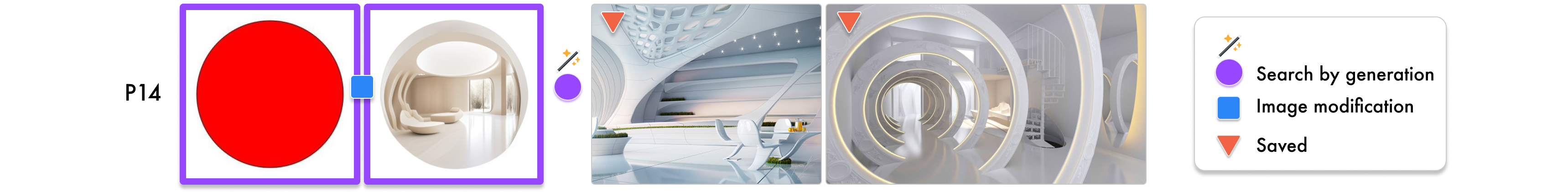}
  \caption{\textbf{P14's \textit{Search by generation} pattern within ~\sysname{}}: P14 searched the basic design layout (the first image on the left) and selected the red circle from the most left-side image. ~\sysname{} provided the ``futuristic'' keyword. Then, he conducted keywords-based editing by inputting "futuristic architecture design" and generated the second image. By doing a \textit{generation by search} with this image, P14 saved several images that were searched from the generation output.}
  \Description{This figure shows P14's generation pattern when he is getting familiar with ~\sysname{}. P14 searched the basic design layout (the first image on the left) and selected the red circle. ~\sysname{} provided the "futuristic" keyword. Then, he conducted keywords-based editing by inputting "futuristic architecture design" and generated the second image. By doing a generation search with this image, P14 saved several images that were searched from the generation output.}
  \label{fig:newGeneration}
\end{figure*}

\sysname{} provided a design process that combined idea search and generation, and as a result, we observed a new design idea was generated through design prototyping-like actions within the search tool. For example, when P14 became familiar with the tool and thought he needed totally different ideas compared to his search history, he proceeded to search for basic design elements (e.g., basic layout) as shown in \autoref{fig:newGeneration}. Then, he tried to edit a more significant part of the image and generate various new ideas. P14 was able to perform design exploration in new directions using the generated results. P14 stated, ``\textit{Even though the generated results were incomplete, the integrated generation results within the search process felt like a collaborator constantly throwing new ideas at me.}'' Besides P14, some participants (P7-P8 and P14) also did additional searches to find design elements (e.g., background image) like a design prototyping process, beyond using the images of existing results. Most of the design tools for search and prototyping are clearly divided so far, as we can check from our formative study. However, by integrating the generation process and search, we found the prototyping and search processes could be interplay by complementing each other. Designers can perform prototyping actions when they need to explore new ideas after a certain level of exploration. Designers can also engage in design exploration actions to further develop raw ideas generated during the prototyping process within ~\sysname{}. We envision future work on generative model-based design support tools that combine search and generation processes, allowing the designers to express their design mode freely.

\subsection{Beyond Keywords Suggestion from Search History}
Through the keyword suggestion in ~\autoref{fig:genInteraction3}, participants were able to define their next direction of exploration while considering the design directions they had explored so far (P1-P2, P4, P7-P8, P11, and P13-P14). P1 and P4 also stated how the user has modified images in ~\sysname{} would be useful information to track the user's intent in addition to the search history information (e.g., search query and saved design description). Furthermore, to this statement, which generation outputs were saved and which ones were used for the search can also serve as valuable clues for \sysname{} to infer the user's visual search intent. By leveraging this information, ~\sysname{} could be further improved to provide search direction feedback on how to modify specific areas of an image and show the expected search results based on the modification.

\subsection{Expanding the Scope: Accommodating Varied Design Intentions}
Our work proposes a method that expresses visual search intent through the outputs of a generation model. Going beyond visual search intent, in future work, our work could be expanded to accommodate various forms of intent that may arise during the design process. For instance, intentions related to modifications during the design prototyping stage or intentions related to creating something new design ideas could be expressed through various modalities of the generation model (e.g., design concept keywords, design style, or design feedback). In addition, by providing the generation model with various types of information from the design process, such as sketches and command sequence patterns used in design tools, we can interpret and reveal the tacit intent of designers. By revealing the vague design intent, future design tools will support a designer by suggesting ideas related to the intent or generating several alternatives. The generative model, offering incomplete but varied modalities, will pose an effective function in future design tools, especially for designers for whom clearly expressing intent poses a challenge.

\subsection{Difference Between the Full Generation-based Exploration and Generation Search-based Exploration}
Several participants (P4, P9, P13, P15-P16) explained that the \sysname{} made them look forward to the search results though they obtained incomplete generation outputs. Specifically, P16 said, ``My expectation of the generated output was lower than compared to my experience with general AI models. Rather, I enjoyed and was curious about the search results based on this incomplete generated output.'' Through participants' responses, the combination of specific task functions would affect the user's expectation on the generated output. Many popular text-to-image models (e.g., DALL·E~\footnote{https://openai.com/dall-e-3}, Firefly~\footnote{https://firefly.adobe.com/}, or Midjourney~\footnote{https://www.midjourney.com}) support a fully generation-based exploration process. For example, Adobe's Firefly provides four outputs for one text prompt, and then the user can select and freely edit the outputs' style, concept, color tone, or composition. Also, the user can generate other similar images based on the most preferred image. However, since the idea exploration process in the early design phase requires a quick process to see various results, full generation-based exploration might be challenging to support this process. Moreover, users may want an output that perfectly fits their intent in this approach. Therefore, based on these findings, future generation-based tools can be designed by considering the level of user expectation on the generation output with the characteristics of the task.

\subsection{Limitations and Possible Approaches for Generation Search}
Even if participants were able to express their desired results through generation visually, there were situations where the related search results were scant. P1 commented, ``\textit{I was reasonably satisfied with the generated output, so I tried searching with them, but I obtained very few search results. It seemed like there was a lack of related content in the dataset, so I explored different design directions.}'' Additionally, P11 mentioned, ``\textit{Since I couldn't see any related search results, I modified the generation results to align them more with what I wanted and then saved those results.}'' When the search results were limited based on the generation output, several potential future research directions can address this issue, including 1) providing guidance for the next generation direction based on the position of the generation results within the dataset, 2) offering additional controllability in the generation process to support the creation of desired results in detail, and 3) modifying the search algorithm. These can all be explored to address this situation effectively.

Besides, ~\sysname{} has the following limitations: First, the modality for selecting the area to modify in an image should allow for more accurate selection. Specifically, the segment-based approach is useful when the object in the image is segmented clearly. However, if the segmented part was too small, the segmentation wasn't clearly divided, or there were some areas including many segments they wanted to edit, the participant wanted to select the area freely. Second, the feature of storing and reusing the modified images, modification process, and their associated search results is required. For example, when a user continuously tries to select the image's background, the system automatically provides selection suggestions based on the editing history. These two limitations were identified based on feedback from the participants in the study. The limitation should be addressed in future work to improve the system further.
\section{Conclusion}
This paper proposes ~\sysname{}, a novel system that integrates generative models into the visual search process. By automatically elaborating on users' queries, \sysname{} can surface concrete search directions when users only have abstract ideas. Also, by generatively modifying existing search results and using these to search for similar images, ~\sysname{} allows users to express what they are looking for more precisely. Our study results revealed the participants felt that they could more accurately express their visual search intent. Through ~\sysname{}, participants felt that they could find more diverse images with more satisfaction. Ultimately, the generation process enhanced the user's creativity in the visual search process by guiding them to new search directions beyond searching for desired images. Although the generation output is unreliable and cannot be fully controlled, ~\sysname{} demonstrated the benefit of leveraging generated outputs as intermediate materials that can represent designers' intents.
\begin{acks}
This work was supported by NAVER-KAIST Hypercreative AI Center.This work was also supported by Institute of Information \& Communications Technology Planning \& Evaluation (IITP) grant funded by the Korea government (MSIT) (No.2021-0-01347,Video Interaction Technologies Using Object-Oriented Video Modeling).
We thank all of our participants for engaging positively in our various studies.
We also thank all of the members of KIXLAB for their helpful discussions and constructive feedback.    
\end{acks}

% Bibliography
\bibliographystyle{ACM-Reference-Format}
\bibliography{references}

% Appendix
\begin{appendices}

\section{Query Concretization Prompt}

\label{appendix:a}
\textbf{system\_prompt:\\} 
\texttt{You are a helpful and creative assistant that can suggest effective search queries to find new and inspiring designs. You return your final answer as a valid JSON object.\\\\}
\noindent
\textbf{template:\\}
\texttt{
    \{prompt1\}\\
    {[Current Search Query]}\\
    \{curr\_query\}\\
    \{prompt2\}\\\\
}
\textbf{prompt1:\\}
\texttt{
    We would like to request you to ideate search queries to help designers explore and find useful reference images. The designer has now entered one text query into the image search system. However, there is currently an unspecified part of this query. If the designer looks for search results with this query, he/she can get too many different search results, so the designer wants to be recommended a more specific search query in the query they enter. These are described below.\\\\
}
\noindent
\textbf{prompt2:}
\texttt{
    \\Please suggest five search queries by following the steps. First, explain the non-specific parts of the current search query and how to specify them. Second, complete the current search query by adding more details to the end regarding color, shape, style, etc. Please add at least three words. Avoid changing the entire meaning of the query, but focus on specifying the unspecified parts in various aspects. \\\\ Return your output as a valid JSON object of the following format: \\ \{"explanation": \\
    <explain how you generate the specified queries in the first and second steps>,\\
    "search\_queries":\\
    {[<list of five suggested queries that designer can use>]}\}\\
}

\section{Editing Keyword Suggestion Prompt}
\label{appendix:b}
\textbf{system\_prompt:\\} 
\texttt{You are a helpful and creative assistant that can suggest effective search queries to find new and inspiring designs. You return your final answer as a valid JSON object.\\\\}
\noindent
\textbf{template:\\}
\texttt{
    \{prompt1\}\\
    {[Description of Current Image]}\\
    \{curr\_image\}\\
    {[Search Query History]}\\
    \{search\_history\}\\
    {[Descriptions of Saved Images]}\\
    \{saved\_images\}\\
    \{prompt2\}\\\\
}
\textbf{prompt1:\\}
\texttt{
    We would like to request you to ideate search terms to help designers explore and find useful reference images. The designer is currently looking at an image. They are trying to think about new search terms that can help them find images that are similar but more inspiring than the current image. The designer has already tried various search queries that were unsuccessful in the past and saved a couple of images to their profile. These are described below.\\\\
}
\noindent
\textbf{prompt2:}
\texttt{
    \\Please suggest several search terms or words. Consider the current image, the previous search history, and the saved images to predict what the designer's intentions may be. You should imagine what type of design the designer is working on and what type of reference images they may be looking for. If the [{Search Query History}] and [{Descriptions of Saved Images}] are empty, just refer only to the [{Description of Current Image}] to predict the designer's intent. Provide a comprehensive explanation about what you imagine the designer's intention to be and the type of reference images that may satisfy or diversify this intent. \\\\
    Then, suggest search terms that can help satisfy the designer's intent. You should suggest search terms that designers can add to their search queries to look for images that satisfy their intentions. As an alternative, also suggest terms that can help diversity the designer's intent. These search terms should be different from the designers current intent and should help them explore other, different types of designs. When suggesting search terms, you should avoid suggesting search terms that are already included in the current image, the search history, or the descriptions of saved images. Ensure that your suggested terms are completely new to the designer. Ensure that you only suggest words and avoid suggesting phrases.\\\\
    Return your output as a valid JSON object of the following format:\\
    \{"explanation": \\
    <explain how you generate the specified queries in the first and second steps>, \\ 
    "aligned\_search\_terms": \\
    {[<list of five suggested words that align with the designer's current intentions>]}, \\ 
    "diversified\_search\_terms": \\
    {[<list of five suggested words that differ from the designer's current intentions>]}\}\\
}
\end{appendices}

\end{document}